\documentclass[a4paper]{llncs}
\usepackage{underscore}

\RequirePackage{amsmath}

\usepackage{listings}
\usepackage{graphicx}
\usepackage{tikz}

\usepackage{xspace}

\usepackage{caption}
\usepackage{makecell}
\usepackage{paralist}
\usepackage{mathtools}
\usepackage{booktabs}
\usepackage{algpseudocode}
\usepackage{version}
\usepackage{multicol}
\includeversion{EXT}
\excludeversion{SHORT}

\begin{EXT}
\end{EXT}

\usepackage[hyphens]{url}
\usepackage[breaklinks=true,pdfborder={0 0 0},bookmarksopen]{hyperref}

\AtBeginDocument{%

}

\lstdefinelanguage{aslanpp}{
  basicstyle= \footnotesize\ttfamily, 
  breakatwhitespace=false,
  mathescape=true,
   numbers=left,
   numberstyle=\tiny,
  morekeywords={specification,channel_model,CCM,ICM,ACM,entity,on,else,import,types,symbols,nonpublic,noninvertible,macros,clauses,equations,body,breakpoints,new,any,where,send,receive,over,retract,if,then,while,select,on,assert,constraints,goals,forall,exists,Actor,for} 
}
\lstMakeShortInline{\|} 

\lstdefinelanguage{pseudo}{
  basicstyle= \footnotesize\ttfamily, 
  breakatwhitespace=false,
  mathescape=true,
   numbers=left,
   numberstyle=\tiny,
  morekeywords={if,else,elseif} 
}
\newcommand{\fix}[2]{{\bf FIX}\footnote{{\bf #1:} #2 }}
\renewcommand{\fix}[2]{}

\newcommand*\circled[1]{\tikz[baseline=(char.base)]{
		\node[shape=circle,draw,inner sep=0.5pt] (char) {#1};}}

\newcommand{\tool}{WAFEx}
\newcommand{\aslanpp}{ASLan++}
\newcommand{\avantssar}{AVANTSSAR Platform}
\newcommand{\clatse}{CL-AtSe}
\newcommand{\wfuzz}{Wfuzz}
\newcommand{\sqlmap}{sqlmap}
\newcommand{\sqli}{SQLi}
\newcommand{\dvwa}{DVWA}
\newcommand{\chained}{Multi-Stage}

\newcommand{\webapplication}{web app}
\newcommand{\Webapplication}{Web app}
\newcommand{\DirectoryTraversal}{DT}
\newcommand{\FileInclusion}{FI}
\newcommand{\ForcedBrowsing}{FB}
\newcommand{\UnrestrictedFileUpload}{UFU}

\begin{document}

\begin{SHORT}
\title{A Formal Approach to Exploiting Multi-Stage Attacks based on File-System Vulnerabilities of Web Applications}
\end{SHORT}

\begin{EXT}
\title{A Formal Approach to Exploiting Multi-Stage Attacks based on File-System Vulnerabilities of Web Applications \\ (Extended Version)}
\end{EXT}

\author{Federico De Meo\inst{1} \and Luca Vigan\`o\inst{2}}
\institute{
Dipartimento di Informatica, Universit\`a degli Studi di Verona, Italy \and
Department of Informatics, King's College London, UK}

\maketitle
\begin{abstract}
\begin{EXT}
Web applications require access to the file-system for many different
tasks. When analyzing the security of a web application, security analysts
should thus consider the impact that file-system operations have on the
security of the whole application. Moreover, the analysis should take into
consideration how file-system vulnerabilities might interact with other
vulnerabilities leading an attacker to breach into the web application.
In this paper, we first propose a classification of file-system
vulnerabilities, and then, based on this classification, we present a formal approach that allows one to exploit file-system vulnerabilities. We give a formal representation of web applications, databases and file-systems, and show how to reason about file-system vulnerabilities. We also show how to 
combine file-system vulnerabilities and SQL-Injection vulnerabilities for the identification of complex, multi-stage attacks. We have developed an automatic tool
that implements our approach and we show its efficiency by discussing several
real-world case studies, which are witness to the fact that our tool can
generate, and exploit, complex attacks that, to the best of our knowledge, no
other state-of-the-art-tool for the security of web applications can find. 
\end{EXT}

\begin{SHORT}
We propose a formal approach that allows one 
to (i) reason about file-system
vulnerabilities of web applications and (ii) combine file-system vulnerabilities
and SQL-Injection vulnerabilities for 
complex, multi-stage
attacks. We have developed an automatic tool that implements our approach and we 
show its efficiency by discussing four real-world case studies, which are
witness to the fact that our tool can generate, and exploit, 
attacks that, to the best of our knowledge, no other 
tool for the security of web applications can
find. 
\end{SHORT}
\end{abstract}

\section{Introduction}
\textbf{\emph{Context and motivations}} 
\begin{EXT}
Every month a large number of novel web applications
(\emph{\webapplication{s}}, for short) are launched that provide useful
functionalities to administer bank accounts, manage personal health
records, sell and buy goods, and so on. These functionalities regularly
attract more users... and attackers! 
Hand-in-hand with the increase of
\webapplication{s} available and features provided,
there has been an exponential increase in the number of
web applications breached to expose
private and sensitive data, with companies and users realizing they were
victims of an attack often only months after it
occurred~\cite{breach:dropbox,breach:tumblr,breach:yahoo}.
%
\end{EXT}
Modern web applications 
\begin{SHORT}(\emph{web apps}, for short) \end{SHORT}
often make intensive use of
functionalities for reading and writing content from the
\emph{\webapplication{}'s file-system} (i.e., the file-system of
the web server that hosts the \webapplication{}). \emph{Reading} from
	and \emph{writing} to the file-system are routine
operations that \webapplication{s} perform for 
different tasks.
For instance, the 
option of dynamically loading resources based on runtime
needs is commonly adopted by developers to structure the \webapplication{}'s
source code for 
stronger reusability. Similarly, 
several \webapplication{s} allow users to
upload (write) content that can be shared with other users or can be available
from a web browser as in a cloud service. Reading and writing functionalities
are offered by most server-side programming
languages for developing \webapplication{s} such as PHP
\cite{phpinclude}, JSP~\cite{jspinclude} or ASP~\cite{aspinclude}. Modern
database APIs also provide a convenient way to interact with the
file-system (e.g., backup or restore functionalities), but they 
also increase the attack surface an attacker could exploit.
Whenever an attacker finds
a way to exploit vulnerabilities that allow him to gain access to the
\webapplication{}'s file-system, the security of the whole
\webapplication{} is put at high risk. Indeed, both 
OWASP~\cite{owasptop10}	and MITRE~\cite{top25}
list vulnerabilities that compromise the file-system among the most common and
dangerous vulnerabilities that afflict the security of modern
software.\footnote{The Top 10 compiled by OWASP is a general classification and
	it does not include a specific category named ``file-system vulnerability'';
	however, ``Injections'', ``Broken Authentication and session Management'',
	``Security misconfiguration'' (just to name a few) can all lead to a
	vulnerability related to the file-system.}

Vulnerability assessment and penetration testing are the two main steps that
security analysts typically undertake when assessing the security of a
\webapplication{} and other computer
systems~\cite{alienvault:vapt,veracore:vapt,sans:vapt}. During a
\emph{vulnerability assessment}, automatic scanning
tools are used to search for common vulnerabilities of the system under analysis (\wfuzz{}~\cite{wfuzz} and DotDotPwn~\cite{ddp} are the main
tools for file-system-related vulnerabilities).
However, it is well known~\cite{johnny} that state-of-the-art scanners do not detect vulnerabilities linked to the logical flaws of \webapplication{s}. 
This means that \emph{even if a vulnerability is found, no tool can link it to
logical flaws leading to the violation of a security property}. The result
of the vulnerability assessment is thus used to perform the second and
more complicated step: during
a \emph{penetration test (pentest)}, the analyst defines
an attack goal and manually attempts to exploit the discovered
vulnerabilities to determine whether the attack goal he defined
can actually be achieved. A pentest is meant to show the real damage on a
specific \webapplication{} resulting from the exploitation of one or more
vulnerabilities. Consider the following example, which is 
simple but also fundamental to understand the motivation for the approach
that we propose. 

Trustwave SpiderLabs found a SQL injection vulnerability in
Joomla!~\cite{joomla}, a popular Content Management System (CMS).
In~\cite{joomlaattack}, Trustwave researchers show
two things: the code vulnerable to SQL injection and how the injection
could have been exploited for obtaining full administrative access. The
description of the vulnerable code clearly highlights an inadequate
filtering of data when executing a SQL query. The description of the
damage resulting from the exploitation of the SQL injection shows that an
attacker might be able to perform a session hijacking by stealing session
values stored as plain-text in the database. The result of this analysis
points out two problems: Joomla! is failing in (1) properly filtering data
used when performing a SQL query and (2) securely storying session values.
Problem (1) could have been identified by vulnerability scanners
(e.g., sqlmap is able to identify the vulnerability), but \emph{no
automatic vulnerability scanner can identify Problem (2) and only a
manual pentesting session is effective}. However, manual pentesting relies
on the security analyst's expertise and skills, making the entire
pentesting phase easily prone to errors. An analyst might underestimate
the impact of a certain vulnerability leaving the entire \webapplication{}
exposed to attackers. \emph{This is why we can't stop at the
identification of a SQL injection or file-system-related vulnerability,
and why we can't address the ensuing attacks with a manual analysis}. Our
approach addresses this by automating the identification of attacks that
exploit such multi-stage vulnerabilities.

\emph{\textbf{Contributions}} 
Our contributions are two-fold. 
First, we formally define 
file-system vulnerabilities and how to exploit them to violate
security properties of \webapplication{s}. A number of formal approaches based on the \emph{Dolev-Yao (DY) attacker model}~\cite{dolev1983} have been developed for the security analysis of \webapplication{s}, e.g., 
\cite{towards,avantssar-tacas,spacite,paper:sqli,csrf,spacios}.
However, 
file-system vulnerabilities of \webapplication{s} have never been taken into
consideration by formal approaches before. In this paper, we define how
\webapplication{s} interact with the file-system and show how the DY model can
be used to exploit file-system vulnerabilities. Moreover, we extend our previous
work on the exploitation of SQL-Injection (\sqli{})
vulnerabilities~\cite{paper:sqli} by showing how to combine file-system
vulnerabilities and \sqli{} vulnerabilities for the identification of complex,
multi-stage attacks commonly identified only by manual analysis during the
pentesting phase.
It is crucial to point out that we do not search for payloads that can be used
to exploit a particular vulnerability, but rather we exploit file-system
vulnerabilities.

Second, to show that our formalization can effectively generate
multi-stage attacks where file-system and \sqli{} vulnerabilities are exploited,
we have developed a prototype tool called \emph{\tool{} (Web Application Formal
	Exploiter, \cite{wafex})} and we discuss here its application to four
real-world case studies. \tool{} can generate, and exploit, complex attacks
that, to the best of our knowledge, no other state-of-the-art-tool for the
security analysis of \webapplication{s} can find. In particular, we show how
\tool{} can automatically generate different
attack traces violating the same security property, a result that only a manual
analysis performed by a pentester can achieve. In each attack trace, multiple
vulnerabilities 
\begin{EXT}(e.g., file-system and \sqli{})\end{EXT} are used in combination.


\emph{\textbf{Organization}} 
\begin{SHORT}
In \autoref{sec:fscat}, we give a classification of file-system-related
vulnerabilities. In \autoref{sec:formalization}, we
provide our formal approach. In \autoref{sec:casestudies}, we describe the
\tool{} tool and discuss its application to four real-world case
studies. In \autoref{sec:concluding}, we draw conclusions and discuss
related work and future work. 
\end{SHORT}
\begin{EXT}
In \autoref{sec:fscat}, we give a classification of file-system
vulnerabilities and describe the advantages an attacker gains in
exploiting such vulnerabilities. In \autoref{sec:formalization}, we
provide our formalization.
In \autoref{sec:casestudies}, we describe the \tool{} tool and discuss its
application to complex real-world case studies. We discuss related work
in \autoref{sec:related}, and we provide conclusions and discuss future
work in \autoref{sec:conclusionsfuturework}. The appendices contain full
details on our formal specifications and case studies.
\end{EXT}


\section{A classification of file-system-related vulnerabilities}
\label{sec:fscat}


To provide a coherent and uniform starting point for
reasoning about file-system-related vulnerabilities,
\begin{EXT}in this section\end{EXT}
we give a classification of the vulnerabilities of
\webapplication{s} that lead to compromise the file-system. 
\begin{SHORT}
The two security properties that we consider are: \emph{authentication} (the
attacker gets unauthorized access to a restricted area) and
\emph{confidentiality} (the attacker gets access to content stored in the
\webapplication{}'s file-system that isn't meant to be publicly available).	
\end{SHORT}
\begin{EXT}
The two security properties that we consider in our formalization are:
\begin{compactitem}
	\item \emph{Authentication bypass}: the attacker gets unauthorized access to a
	restricted area.
	\item \emph{Confidentiality}: the attacker gets access to
	the content stored in the \webapplication{}'s file-system that is not meant
	to be publicly available.
\end{compactitem}
\end{EXT}
We have identified five vulnerability categories, which we describe below,
focusing on the main details of the attacks that are relevant for our
formalization.

\emph{\textbf{(1) Directory Traversal (DT)}} (a.k.a.~\emph{Path Traversal})
Operations on files (reading and writing) performed by a \webapplication{} are
intended to occur in the \emph{root directory} of the \webapplication{}, a
restricted directory where the \webapplication{} actually resides. 
A \emph{DT vulnerability} refers to a
lack of authorization controls when an attacker attempts to access a location
that is intended to identify a file or directory stored in a restricted area
outside the \webapplication{}'s root directory.
Whenever the access permissions of a \webapplication{} are not restricted in
such a way that they only allow users to access authorized files, an attacker
might be able to craft a payload that allows him to access restricted files
located outside the \webapplication{}'s root directory.
\begin{EXT}
\DirectoryTraversal{} payloads make use of special characters such as the
double dots ``\texttt{..}'' and the forward slash ``\texttt{/}'' separator,
which, when combined, allow the attacker to specify arbitrary locations that
can escape outside the root directory of a \webapplication{}.
\DirectoryTraversal{} attacks can be further divided in \emph{Relative
	\DirectoryTraversal{}} and \emph{Absolute \DirectoryTraversal{}}, depending
on whether the payload refers to a relative or an absolute path.  Since a
\DirectoryTraversal{} vulnerability refers to a lack of authorization
permissions, to actually exploit it, it is necessary for an attacker to find an
entry point that allows him to send input to the \webapplication{} that is then
used to create a file-location string. This means that a \DirectoryTraversal{}
vulnerability is always exploited in combination with another vulnerability
that provides such an entry point to the attacker.  For example, imagine that
\texttt{index.php?load=file} refers to a web page \texttt{index.php} that
dynamically loads the file specified by the value of \texttt{load}. An attacker
might modify this value and use it as input vector to exploit a
\DirectoryTraversal{} vulnerability.
\end{EXT}

\emph{\textbf{(2) SQL-Injection (\sqli{})}}
\begin{EXT}
\Webapplication{s} make use of a \emph{Database Management System (DBMS)} in
order to store data. This allows for functionalities such as blog posting,
forum discussions, etc. Querying a database is performed using the Structured
Query Language SQL and whenever a query is created using user-supplied data,
\emph{\sqli{} attacks} could be
possible~\cite{paper:sqli,halfond06mar,owasp:sqli}.
\end{EXT}
Most modern DMBSs provide APIs that extend SQL's expressiveness by
allowing SQL code to access a \webapplication{}'s file-system for reading
and writing purposes. Reading APIs allow developers to produce code that
retrieves content stored in the \webapplication{}'s file-system and loads
it in the database. 
\begin{EXT}This is particularly convenient when a
\webapplication{} needs to load bulks of data into the database, e.g., as
part of an initialization or restoring process.
\end{EXT}Writing APIs allow
developers to produce code that saves content from the database to the
\webapplication{}'s file-system. 
\begin{EXT}This is particularly 
convenient for features such as backup or upload functionalities.  When an
attacker finds an \sqli{} entry point, he can inject arbitrary SQL syntax that
modifies the behavior of the original query.\end{EXT}
Attackers mainly exploit \sqli{} to bypass authentication mechanisms or to
extract data from the database, but, as there is no limit on the SQL syntax
that could be injected, it is also possible to exploit reading and writing APIs
to access the underlying file-system~\cite{Damele2009}.

\begin{EXT}
As an example, consider the MySQL DBMS~\cite{MySQL}, which has the
built-in API \texttt{LOAD\_FILE()} for reading text or binary files from
the file-system~\cite{mysqlloadfile}. To execute \texttt{LOAD\_FILE()},
the user executing the query needs to have the \texttt{FILE} privileges,
which give the DBMS permission to read and write files on the server. The
file accessed by the attacker must be owned by the user that started MySQL
or be readable by all users. Similarly, MySQL provides APIs for writing to
the file-system with the \texttt{SELECT ... INTO} statement, which enables
the result of a \texttt{SELECT} query to be written to a file. For
instance, in presence of a \texttt{UNION} query-based \sqli{}, an attacker
might inject the payload \texttt{1 UNION ALL SELECT
1,LOAD\_FILE('/etc/passwd'),3,4 FROM mysql.user--} that will give him
reading access to the file \texttt{/etc/passwd}.
\end{EXT}


\emph{\textbf{(3) File Inclusion (FI)}}
All programing languages for the development of \webapplication{s} support
functionalities for structuring code into separate files so that the same
code can be reused at runtime by dynamically including files whenever
required. A \emph{\FileInclusion{} vulnerability} refers to a lack of
proper sanitization of user-supplied data during the creation of a file
location string that will be used to locate a resource meant to be
included in a web page. When the file location depends on user-supplied
data, an attacker can exploit it and force the inclusion of files
different from the ones intended by the developers. FI
might allow an attacker to access arbitrary resources stored on the
file-system and to execute code.
\begin{EXT}
\FileInclusion{} attacks can be further divided in \emph{Local
\FileInclusion{}} and \emph{Remote \FileInclusion{}}, which force the
inclusion of files stored locally or remotely, respectively.



As an example, consider the PHP code in \autoref{lst:ptht}. Line (2) gets
a user-supplied parameter \texttt{\$\_GET['user']} and stores it into the
variable \texttt{\$username} that is then used to create a file location
(3), which is in turn used to include a resource in the current page (4).
If, as the \texttt{user}'s value, an attacker injects the payload
\texttt{.htaccess}, he might access the \texttt{.htaccess} file, which
allows one to make configuration changes on a per-directory
basis~\cite{apache:htaccess}.

\begin{lstlisting}[basicstyle=\footnotesize\ttfamily,numbers=left,xleftmargin=2em,language=php,mathescape=false,caption=PHP code vulnerable to \FileInclusion{}.,label=lst:ptht,showstringspaces=false, breaklines=true]
$username = $_GET['user'];
$filepathname = "/var/www/html/".$username;
include $filepathname;
\end{lstlisting}

\FileInclusion{} can be combined with the \DirectoryTraversal{} vulnerability,
allowing an attacker to gain access to resources stored on the
\webapplication{}'s file-system but outside the \webapplication{}'s root
folder. 
Consider again \autoref{lst:ptht} and suppose that the web server hosting the
\webapplication{} is a unix server. The attacker might then inject a malicious
payload such as \texttt{../../../../etc/passwd}, where \texttt{/etc/passwd} is
the common location pointing to a text-based database listing the users of the
system. Assuming that the root directory of the \webapplication{} is located at
\texttt{/var/www/html/site/}, the PHP code will try to include the file
\texttt{/var/www/html/site/../../../../etc/passwd}, and the path is translated
into \texttt{/etc/passwd}, forcing the \webapplication{} to include the file
and thus giving the attacker access to the list of users for the web server.
\end{EXT}

\emph{\textbf{(4) Forced Browsing (\ForcedBrowsing{})}}
(a.k.a.~\emph{Direct Request}) refers to a lack of authorization controls when
a resource is directly accessed via URLs. This lack of authorization might
allow an attacker to enumerate and access resources that are not referenced by
the \webapplication{} (thus not directly displayed to the users through the
\webapplication{}) or that are intended to be accessed only as a result of
previous HTTP requests. By making an appropriate HTTP request, an attacker
could access resources with a direct request rather than by following the
intended path.  The lack of authorization controls comes from the erroneous
assumption that resources can only be reached through a given navigation path.
This misassumption leads developers to implementing authorization mechanisms
only at certain points along the way for accessing a resource, leaving no
controls when a resource is directly accessed.

\begin{EXT}
As an example, consider a \webapplication{} where the URL
\texttt{http://vuln.com/} \texttt{admin/index.php} points to the
login page and \texttt{http://vuln.com/admin/admin. php}
points to the administration page accessible once the login has succeeded.
When \ForcedBrowsing{} is possible, an attacker might be able to directly
access the administration page by requesting the URL
\texttt{http://vuln.com/admin/admin.php} without first logging in.
If the \texttt{admin.php} page does not verify whether the request is made
by an authorized user, the attacker has skipped the login process provided
by \texttt{http://vuln.com/admin/index.php}.

Another example is a development environment that is supposed to be
accessible only by developers. 
Developers usually erroneously assume that since the development environment is
not directly accessible from the main website, users have no means to access
this area, so an attacker might try to guess the name of the development
environment (e.g., \texttt{http://vuln.com/dev/}) and increase the chance of
having unauthorized access to the \webapplication{}.
\end{EXT}

\emph{\textbf{(5) Unrestricted File Upload (UFU)}}
A widespread feature provided by \webapplication{s} is the possibility of
uploading files that will be stored on the \webapplication{}'s file-system. An
\emph{\UnrestrictedFileUpload{} vulnerability} refers to a lack of proper file
sanitization when a \webapplication{} allows for uploading files.  The
consequences can vary, ranging from complete takeover with remote arbitrary
code execution to simple defacement, where the attacker is able to modify the
content shown to users by the \webapplication{}.

\begin{EXT}
As an example, consider a \webapplication{} that allows users to upload images
(e.g., an avatar to customize the user profile).  \autoref{lst:uphtml} gives
the HTML code of the file upload form, \autoref{lst:upphp} gives the PHP code
that performs the upload but without carrying out any control over the file
being uploaded, paving the way to \UnrestrictedFileUpload{} attacks.
\begin{lstlisting}[basicstyle=\footnotesize\ttfamily,language=html,mathescape=false,caption=HTML code that shows a file upload form.,
				label=lst:uphtml,showstringspaces=false, breaklines=true]
<form enctype="multipart/form-data" action="uploader.php" method="POST">
Choose a file to upload: <input name="uploadedfile" type="file" /><br />
<input type="submit" value="Upload File" /> </form>
\end{lstlisting}

\begin{lstlisting}[basicstyle=\footnotesize\ttfamily,language=php,mathescape=false,caption=PHP code for uploading a file.,
				label=lst:upphp,showstringspaces=false, breaklines=true]
$path="uploads/";
// where the file will be saved
$target_path = $target_path.basename($_FILES['uploadedfile']['name']);
// move uploaded file from tmp directory to target path
if(move_uploaded_file($_FILES['uploadedfile']['tmp_name'],$target_path)){
  echo "file " . basename($_FILES['uploadedfile']['name']) . " uploaded!";
}else { echo "there was an error uploading the file"; }
\end{lstlisting}

\UnrestrictedFileUpload{} can also be used in combination with
\DirectoryTraversal{}, allowing the attacker to overwrite arbitrary files
stored in the web server. However, modern \webapplication{} programming
languages (like PHP) perform a sanitization on the uploaded file path
string by removing any special characters such as ``\texttt{..}'' and
``\texttt{/}'' used to change the current path, making the exploitation of
a \DirectoryTraversal{} less likely to happen.
\end{EXT}

\section{A formalization to reason about file-system vulnerabilities}
\label{sec:formalization}

We will now describe how we formally represent the behavior of a
file-system and of a \webapplication{} that interacts with it. We also
show how our formalization allows the DY attacker to successfully exploit
file-system vulnerabilities. In our formalization, we used ASLan++, the
formal specification language of the AVANTSSAR
Platform~\cite{avantssar-tacas}, but in fact our approach is general and
for the sake of readability we give here pseudo-code rather than ASLan++.
\begin{EXT}
(see the appendix for the full ASLan++ code).
\end{EXT}

Our approach doesn't search for payloads that can be used to exploit
file-system vulnerabilities, but rather analyzes the security of
\webapplication{s} by exploiting vulnerabilities that lead attackers to have
unauthorized access to the file-system. To deal with such 
vulnerabilities, we need to represent 
the \emph{behavior of the}
\begin{compactitem}
\item[(i)] \emph{\webapplication{}}, which defines the
	interaction with
	\begin{EXT}the\end{EXT}
	client, \begin{EXT}the\end{EXT}
	file-system and \begin{EXT}the\end{EXT} database, 
\item[(ii)] \emph{file-system}, which interacts with the
	\webapplication{} and the database for reading and writing content, 
\item[(iii)] \emph{database}, which can also interact with the file-system,  
\item[(iv)] \emph{attacker}, who interacts only with the \webapplication{}. 
\end{compactitem}

We do not formalize the behavior of honest clients since we assume the DY
attacker to be the only agent to communicate with the \webapplication{},
i.e., we are only interested in dishonest interactions. This is because
the exploitation of file-system vulnerabilities doesn't require
interaction with the honest users.

\begin{figure}[t]
\begin{center}
\includegraphics[scale=0.58]{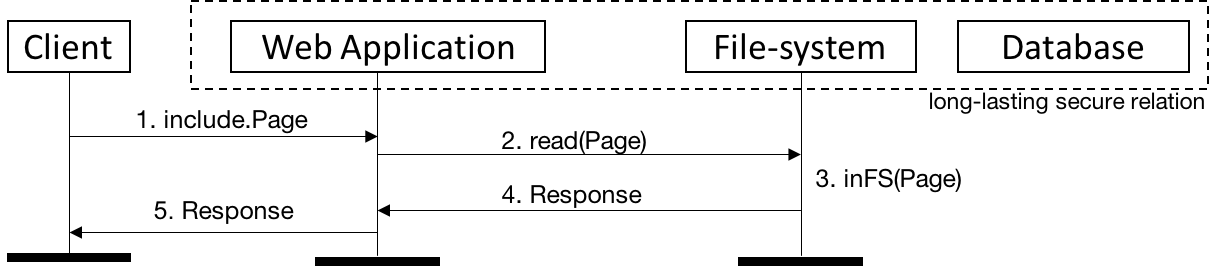}
\caption{MSC of a \FileInclusion{} vulnerability.}
\label{fig:msc}
\end{center}
\end{figure}

To explain how our formalization works, we will use a simple
\FileInclusion{} example depicted in \autoref{fig:msc} as a \emph{Message
Sequence Chart (MSC)} in which there are three entities: client,
\webapplication{} and file-system (the fourth entity, database, does not
send messages and will be discussed later). In this example and, in
general, in our formalization, we assume the \webapplication{} and the
file-system (and the database) to have a long-lasting secure relation,
i.e., no attacker can read or modify the communication between
them.\footnote{We assume the communication with the file-system to be
secure since the file-system actually is not a real network node, and thus
no attacker can put himself between the communication with the
file-system, i.e., man-in-the-middle attacks are not possible.} 
Moreover, as is standard,
constants begin with a lower case character (e.g., \texttt{filePath}),
variables with an upper case one (e.g., \texttt{Page}).\fix{F}{questo Page va a capo}

In step
(1), the client sends to the \webapplication{} a message containing the
variable \texttt{Page}, representing the page to be included. In step (2), the
\webapplication{} performs a read operation by issuing a
\texttt{read(Page)} request. In step (3), the file-system checks if the
requested page points to an existing file and, in step (4), it replies to the
\webapplication{} with a variable \texttt{Response} that represents the
result of the read operation. In step (5), the \webapplication{} forwards the
response to the client.

\begin{definition} 
\emph{Messages} consist of \emph{variables} $V$, \emph{constants} $c$,
\emph{concatenation} $M.M$, \emph{function application} $f(M)$ of
uninterpreted function symbols $f$ to messages $M$, and \emph{encryption}
$\{M\}_M$ of messages with public, private or symmetric keys that are
themselves messages.\footnote{In this paper, we need not distinguish
between different kinds of encrypted messages, but we could 
do it by following standard practice. Here we don't even need to consider
explicitly encrypted messages, but we add them for completeness.} 
We define that $M_1$ is a \emph{submessage} of $M_2$ as is standard (e.g.,
$M_1$ is a submessage of $M_1.M_3$, of $f(M_1)$ and of $\{M_1\}_{M_4}$) and,
abusing notation, we then write $M_1 \in M_2$. 
\end{definition}


\subsection{The DY attacker as a web attacker}
The DY model~\cite{dolev1983} defines an attacker that has
total control over the network but cannot break cryptography: he can
intercept messages and decrypt them if he holds the corresponding keys for
decryption, he can generate new messages from his knowledge and send
messages under any agent name. Message generation and analysis are
formalized by derivation rules, expressing how the attacker can derive
messages from his knowledge in order to use them for performing attacks.

Suppose that we want to search for a \FileInclusion{} vulnerability
possibly leading the attacker to access resources stored outside the root
folder of the \webapplication{} (\DirectoryTraversal{}). As described in
\autoref{sec:fscat}, the attacker can try to access resources by injecting
a payload used for accessing a file that will be included in the current
page. However, it is important to point out two fundamental aspects of
our work: (1) as stated, we are not interested in generating the payloads that will
exploit vulnerabilities, but rather we want to represent that a
vulnerability can be exploited and what happens when it is exploited, and
(2) we want to avoid state-space explosion by making the models as simple
as possible. We have thus introduced the constant \texttt{fsi} that
represents any and all payloads for exploiting file-system
vulnerabilities (e.g., \texttt{../../../} for
	\DirectoryTraversal{}). By
using  \texttt{fsi}  and the definition of the file-system
entity (\autoref{sec:fs}), we allow the DY attacker to deal with file-system
vulnerabilities.

\subsection{File-system}
\label{sec:fs}

We will now give a formalization of the file-system entity that can be used in
any specification when searching for attacks related to file-system
vulnerabilities of \webapplication{s}. As depicted in~\autoref{fig:msc}, the
file-system can be seen as a network node always actively listening for
incoming connections, and the \webapplication{} sends reading and writing
requests to the file-system.

Our formalization aims to abstract away as many concrete details as possible,
while still being able to represent the exploitation of file-system
vulnerabilities, and so we do not represent the file-system structure but
rather formalize messages sent and received along with reading and writing
behavior. This allows us to give a compact formalization so as to avoid
state-space explosion problems 
when carrying out the analysis with the model checker.

\emph{\textbf{Incoming messages}}
As incoming messages for the file-system entity, we consider only reading
and writing requests, for which we use the uninterpreted functions
\texttt{readFile()} and \texttt{writeFile()}, both taking
a generic variable \texttt{Filepath} that represents a file location in the file-system.

\emph{\textbf{Reading and writing behavior}}
To exploit vulnerabilities related to reading and writing operations, we
need to represent the available files in addition to the behavior of
the two operations. We represent the existence of a file by means of the
predicate \texttt{inFS()}, i.e., \texttt{inFS(filePath)} means that the
file represented by the constant \texttt{filePath} is stored in the
file-system.

When the file-system entity receives a reading request
\texttt{readFile(filePath)}, it checks, by means of \texttt{inFS()}, whether
the file exists: if so, then it answers to the request with
\texttt{file(filePath)}, i.e., wrapping the file being read with the
uninterpreted function \texttt{file()}, else the constant \texttt{no\_file} is
returned.\footnote{We don't need to consider access control policies/models as
	they are external to the \webapplication{}. Hence, we assume that every file
	that is in the file-system can always be read and that every writing
	operation will always succeed.\label{footnote-ac}}


When the file-system entity receives a writing request
\texttt{writeFile(file)}, it uses \texttt{inFS()} to mark the file as part of
the file-system but it need not return a result to the \webapplication{} (as
explained in \autoref{footnote-ac}).

To represent an attacker's attempt to access files, we include the Horn clause
\begin{lstlisting}[numbers=none] 
  fs_hc_evil(M):inFS(fsi.M) 
\end{lstlisting} 
that states that the predicate \texttt{inFS()} holds for a message whenever it
is of the form \texttt{fsi}.$M$, i.e., a message that is a concatenation
including the constant \texttt{fsi} that represents a payload to exploit
file-system vulnerabilities. More specifically, this states that the attacker
has injected a malicious payload \texttt{fsi} into the parameters (expressed as
a variable) $M$. In case of a \DirectoryTraversal{} attack, one may think of
\texttt{fsi} as the \texttt{../../../} payload that escapes from the
\webapplication{}'s root folder.

%

\emph{\textbf{The specification}}
Summarizing, the pseudo-code representing the file-system behavior
is in~\autoref{lst:filesystemEntity}
\begin{SHORT}
(the full ASLan++ specification is in~\fix{F}{Extended version}).
\end{SHORT}
\begin{EXT}
(the full ASLan++ specification is in~\autoref{appendix:implementationaslanpp}).
\end{EXT}
We represent the file-system as a network node always actively listening
for incoming messages. More specifically, we define the file-system by two
mutually exclusive branches of an if-elseif statement: in the guard
in line (1) the file-system receives (expressed in Alice\&Bob notation)
a reading request and in (4) it receives a writing request.
For the reading request, the file-system verifies the
existence of the file (2): if the file exists, then the file-system
returns the variable \texttt{Filepath} wrapped with the uninterpreted
function \texttt{file()} (2), else the constant \texttt{no\_file} is returned
instead (3, where \texttt{!} formalizes 
negation). As we assume that writing operations always succeed, when a
writing request is received, the file-system marks the new file as ``existing''
by means of \texttt{inFS()} and need not return (4).

\begin{lstlisting}[language=pseudo,caption={Pseudo-code representing the
				behavior of the file-system; we write \texttt{FS} for the
				file-system and \texttt{Entity} as a general entity (either \webapplication{} or database).},
label=lst:filesystemEntity]
if(Entity -> FS: readFile(Filepath)){
 if(inFS(Filepath)){ 
  FS -> Entity : file(Filepath); 
 } elseif(!(inFS(Filepath))){
  FS -> Entity : no_file;
} elseif(Entity -> FS: writeFile(Filepath)){ 
  inFS(Filepath); 
}
\end{lstlisting}

\subsection{Database}
\label{sec:db}

To cover all file-system vulnerabilities (\autoref{sec:fscat}),
we need to formalize a database that can interact with the file-system. We
can 
adapt the basic formalization we gave in~\cite{paper:sqli} for
the case of \sqli{} vulnerabilities 
by including 
interaction with the file-system (and thus to also be able to cover file-system
access through \sqli{} vulnerabilities, see 
\autoref{sec:fscat}).
We can see the database, like the file-system, as a network node that interacts
with the \webapplication{} and the file-system.
The idea behind 
the extension is to make the database able to perform a reading
or writing request to the file-system whenever a query is valid. We also 
modified how sanitized queries are handled by removing the sanitization
function \texttt{sanitizedQuery()} from the database specification
of~\cite{paper:sqli} and introducing a new uninterpreted function
\texttt{sanitized()} that represents a general sanitization function
(see~\autoref{sec:webapp} for further details). 
\begin{EXT}
For brevity, we give in~\autoref{appendix:db} a description of
the database behavior as given in~\cite{paper:sqli} while in this section we
focus on the extension.
\end{EXT}

The pseudo-code of the extension is shown in
\autoref{lst:databaseEntityExt}.
\begin{SHORT}
We focus only on the new part and refer to~\cite{paper:sqli} for the main behavior. 
\end{SHORT}
The database entity is still represented by an if-elseif statement. The main if
branch (1) handles sanitized queries, represented with the new sanitization
function, whereas the second branch (3) handles raw queries.  Within the raw
query branch, we have defined two additional behaviors. The
first new branch 
performs a read operation on the file-system (5) and,
if the file-system sends back a file (6), the database wraps the answer
from the file-system with 
\texttt{tuple()} and sends it back to the \webapplication{} (6). The
second new branch (7) handles writing operations performed by the
database, for which the answer to the \webapplication{} will be a message
of the form \texttt{tuple(newFile(filePath))}, where \texttt{newFile()} is
an uninterpreted function stating that a file has been written as a result
of a SQL query.

\begin{lstlisting}[caption={Pseudo-code representing
the behavior of the extended database.},label={lst:databaseEntityExt}]
if(WebApp -> DB: query(sanitized(SQLquery))){
 if(SQLquery == tuple(*)){ 
  DB -> WebApp: no_tuple; 
}} elseif(WebApp -> DB: query(SQLquery)){
 if(inDB(sqlquery)){ 
  DB -> WebApp: tuple(SQLquery); 
 } elseif(inDB(SQLquery)){ 
   DB -> FS: read(SQLquery);
   if(FS -> DB: file(SQLquery)){ 
    DB -> WebApp: tuple(file(SQLquery));
}} elseif(inDB(SQLquery)){ DB -> FS: write(SQLquery);
  DB -> WebApp: tuple(new_file(SQLquery));
} elseif(!(inDB(SQLquery))){ 
  DB -> WebApp: no_tuple; 
}}
\end{lstlisting}

\subsection{Web application}
\label{sec:webapp}

The \webapplication{} is another node of the network that can send and
receive messages. In our formalization, the \webapplication{} can
communicate with client, file-system and database. In~\cite{paper:sqli} we
also provided some guidelines on how to represent \webapplication{s},
however, they were limited to basic interaction with the database. In this
paper we provide extended guidelines that also take the file-system into
consideration.



The file-system and the database entities don't depend on the \webapplication{}
and thus we can reuse them in every model.  The \webapplication{} formalization
does depend on the scenario being modeled, but we give a series of guidelines
on how to represent the \webapplication{}'s behavior for testing the
interaction with the file-system and the database.

\emph{\textbf{If statements}}
HTTP 
is a stateless protocol, which means that each pair
request-response is considered as an independent transaction that is not
related to any previous request-response. We use if statements to define
that a \webapplication{} can answer different requests without following a
specific sequence of messages, thus representing the stateless nature of
HTTP. The \webapplication{}'s model is thus a sequence of if statements
defining all the requests the \webapplication{} can handle.


\emph{\textbf{Client communication}}
A general HTTP request (and response) header comprises different fields
that are needed for the message to be processed by the browser. In our
formalization, we don't need to represent all the fields of a real request
(or response) header as they are not relevant to the analysis, and thus we
limit to: a variable representing the sender, a variable representing
the receiver, and a concatenation of constants and variables representing
the message. 

In case of a request, the message 
would be represented by the HTTP query string containing parameters and values. For example, a request to the URL
\texttt{http://example.com/index.php?page=menu.php} can be represented as
\texttt{Client -> WebApp : index.Page}, where \texttt{index} is a constant
representing a web page and \texttt{Page} is a variable representing an
HTTP query value.
We proceed similarly to formalize a response from the \webapplication{} to
the client. We only represent the details needed for the analysis:
\begin{SHORT} 
a constant representing the returned page  (e.g., \texttt{admin}, \texttt{dashboard}, ...),
the function \texttt{file()} when the \webapplication{} performs a reading operation on the file-system, and
the function \texttt{tuple()} that 
is returned only if the executed query is \texttt{SELECT}, \texttt{UPDATE} or
\texttt{DELETE} (see~\cite{paper:sqli} for further details).
\end{SHORT}
\begin{EXT}
\begin{compactitem} 
\item a constant representing the returned page  (e.g., \texttt{dashboard},
	\texttt{admin} etc),
\item the function \texttt{file()} when the \webapplication{} performs a
	reading operation on the file-system, and
\item the function \texttt{tuple()} that, following~\cite{paper:sqli}, is
	returned only if the executed query is \texttt{SELECT}, \texttt{UPDATE} or
	\texttt{DELETE}.
\end{compactitem} 
\end{EXT}
For example, the response \texttt{WebApp -> Client :
dashboard.file(fsi)} can be used to represent the result of
a successful request where the client receives the \texttt{dashboard}
page. The message \texttt{file(fsi)} is returned to express that a
file was retrieved from the file-system.

\emph{\textbf{File-system and database communication}}
As already stated in \autoref{sec:fs}, whenever the \webapplication{} has
to read content from the file-system, it sends a \texttt{readFile()}
request and whenever it has to write to the file-system, it sends a
\texttt{writeFile()} request. When the \webapplication{} has to perform a
SQL query on the database, it sends a \texttt{query()} request to the
database (see~\autoref{sec:db}).
To allow the \webapplication{} to represent sanitized input, we introduced an
uninterpreted function \texttt{sanitized()} that
allows the modeler to ``switch on'' or ``switch off'' the possibility of
exploiting a vulnerability either of the file-system or of the database,
letting the model-checker analyze the \webapplication{} for possible
attacks.
The \webapplication{} has to check the response coming from the
file-system or the database in order to behave properly. For example, if a
file is being read, the \webapplication{} has to check that the
file-system is answering with the uninterpreted function \texttt{file()}
before proceeding further.

\emph{\textbf{Remote code execution}}
Our formalization can represent scenarios where the attacker is able to
write arbitrary files into the \webapplication{}'s file-system leading to
arbitrary remote code execution. As described earlier, 
a \webapplication{} model is a sequence of if statements defining the
requests the \webapplication{} responds to. The possibility of uploading a
file that leads to code execution can be seen as a way of increasing the
number of requests the \webapplication{} responds to. We then include into
the model of the \webapplication{} a series of predefined if branches
representing the behavior of common malicious code an attacker might try
to upload. We define that these malicious if branches can be used by the
attacker only if the file exists in the file-system (i.e., \texttt{inFS()}
is valid for that file). This will ensure that the attacker finds a way of
writing the malicious file before actually using it.

\emph{\textbf{Sessions}}
As already mentioned, HTTP is a stateless protocol. In order for the user
to experience a stateful interaction with a \webapplication{} (e.g., the
\webapplication{} recognizes when a user is logged-in when he changes
page), developers make use of sessions. A \emph{session} allows a
\webapplication{} to store information into a memory area in order to have
it accessible across multiple pages. When a request is made to a web page
that creates a session, the web page allocates a memory area and assigns
to that area a session identifier. The same session identifier is sent
back to the client (generally as a cookie value) within the response for
that request. When a cookie is received, a web browser automatically sends
it back to the \webapplication{} when a new request is made. The
\webapplication{} receives the session identifier and uses it to retrieve
the information stored within the associated memory area.

In order to represent sessions, we introduced the predicate
\texttt{sessionValue()} to state that a variable is a session value.
Whenever a request is made to a page that creates a session value, a new
variable is created, marked as a session value and returned to the client.
Whenever a page requires a session prior to performing any further step,
the page needs to verify that one of the variables sent to the
\webapplication{} is indeed a session value.

\emph{\textbf{Taking stock}}
\autoref{lst:includeExample} can now finally formalize the \FileInclusion{} example of \autoref{fig:msc}.
The \webapplication{} accepts a request for \texttt{include.Page}, where \texttt{include} is a constant representing the web page being requested and \texttt{Page} is a variable representing the page to include (1).
The \webapplication{} sends a reading request with the page received by the
client (1). The file-system checks if the file is stored in the file-system and
then sends a response to the \webapplication{} (2): if the response is of the
form \texttt{file(Page)} (3), then the \webapplication{} sends back to the
client \texttt{include} (representing an included web page) along with
\texttt{file(Page)} (representing the content of the included file), else it
sends \texttt{include} without further details (4).

\begin{lstlisting}[caption={Pseudo-code representing the behavior of \FileInclusion{} example of \autoref{fig:msc}.}, label={lst:includeExample}]
if(Client -> WebApp: include.Page){ 
  WebApp -> FS : read(Page);
  FS -> WebApp: Response;
  if(Response == file(Page)){ 
   WebApp -> Client: include.file(Page); 
  }else{ 
   WebApp -> Client: include; 
  }}
\end{lstlisting}

\subsection{Goals}

The last component of the formalization is the goal (or security property) that
should be verified. As we discussed in \autoref{sec:fscat}, we are not
interested in finding file-system vulnerabilities but rather we want to exploit
them. In particular, we define 
security properties related to authentication bypass and confidentiality
breach, which, respectively, express that the attacker can access some part of
the \webapplication{} that should be protected with some sort of authorization
mechanisms, or obtain information that is ``leaked'' from the \webapplication{}
(such ``leakage'' can happen from either the file-system or the database).


We use the LTL ``globally'' operator \texttt{[]}, which defines that a formula
has to hold on the entire subsequent temporal path, and the \texttt{iknowledge}
predicate, which represents the knowledge of the attacker. We can then
represent authentication goals by stating that the attacker will never have
access to some specific page, and confidentiality goals by stating that the
attacker will never increase his knowledge with parts coming from the
file-system or the database, i.e., \texttt{file()}.  The confidentiality goal for the \FileInclusion{} example in~\autoref{fig:msc} is shown in \autoref{lst:goals}, stating that the attacker will never know something of the form \texttt{file()}.

\begin{lstlisting}[language=pseudo,numbers=none,caption={Confidentiality goal for the \FileInclusion{} example in~\autoref{fig:msc}.},
		label={lst:goals}]
[](!(iknowledge(file(*))))
\end{lstlisting}

\section{Our tool \tool{} and its application to case studies}
\label{sec:casestudies}

In this section, we show how our formalization can be used effectively for
representing and testing attacks involving the exploit of file-system
vulnerabilities. We have developed a prototype tool, called \emph{\tool{}},
that shows how the \emph{abstract attack trace (AAT)} generated from our models
can be concretized and tested over the implementations of the real
\webapplication{s}. We have tested \tool{} on \emph{Damn Vulnerable Web
	Application (\dvwa{})}~\cite{dvwa} and on \emph{\chained{}}, a
\webapplication{} we wrote for security testing and freely available
at~\cite{wafex}. \dvwa{} is a vulnerable \webapplication{} that provides an
environment in which security analysts can test their skills and tools. \dvwa{}
is divided in examples implementing web pages vulnerable to the most common
\webapplication{} vulnerabilities. We selected three relevant exercises from
\dvwa{}: \FileInclusion{}, \UnrestrictedFileUpload{} and \sqli{}.  \tool{} was
able to identify the intended vulnerability on all the case studies and was
also able to identify an unintended vulnerability of \sqli{} for file reading
in the \sqli{} exercise of \dvwa{}.

In the remainder of this section, we will describe \tool{} and present the
\chained{} case study that shows how file-system vulnerabilities and \sqli{}
can be combined for the generation of multi-stage attack traces on the same
\webapplication{}. 

Our pool of case studies might look small and trivial at first, but
it is worth noting that \dvwa{} is a state-of-the-art testing environment
used for teaching the security of \webapplication{s} to pentesters and the case
studies represent real scenarios that might be implemented by many
\webapplication{s}. Moreover, ethical aspects prevented us from blindly
executing \tool{} on the Internet since our implementation makes use of
brute-forcing tools such as \wfuzz{} and \sqlmap{}, whose unauthorized usage
must be approved by the owner of the \webapplication{}.\fix{Luca}{Ma l'uso di
	ZAP \`e etico?\textbf{F} dipende, se usato come PROXY s\`i, se usato come
	scanner senza autorizzazione, no}
Finally, we didn't test the latest release of any
free CMSs as that would require a first phase of vulnerability
assessment, which is out of the scope of this work.


We give here only the pseudo-code
specification of the \webapplication{s} and the goals; full models in
pseudo-code and \aslanpp{} can be found
\begin{SHORT}
	\fix{F}{Extended version}
\end{SHORT}
\begin{EXT}
in~\autoref{appendix:casestudies} and~\autoref{appendix:implementationaslanpp},
respectively.
\end{EXT}

\subsection{\tool{}: a Web Application Formal Exploiter}
\tool{} takes in input an \aslanpp{} specification
together with a concretization file that contains information such as the
real URL of the \webapplication{} and the name of the real parameters used
for making requests\begin{SHORT}.\end{SHORT}
\begin{EXT}(see \autoref{fig:workflow}
	in \autoref{appendix:workflow} for a detailed description of the workflow of
our approach and tool). 
\end{EXT}
We have chosen \aslanpp{} so to 
apply the
model-checkers of the \avantssar{}~\cite{avantssar-tacas} 
(in particular, \clatse{}), 
but our approach is general and could be 
used with other specification languages and/or other
reasoners implementing the DY attacker model.


To aid the security analyst in the model creation, we have written a Python
script that allows him to first use the Burp proxy~\cite{burp} to record a
trace of HTTP requests/responses generated by interacting with the
\webapplication{}, and then use our script to convert this trace into an
ASLan++ model. The analyst has to specify the behavior for the HTTP requests
and the security goal he wants to test.

\begin{EXT}
%
By using the \aslanpp{} translator~\cite{avantssar-tacas}, \tool{}
generates a transition system 
that is given in input to the model-checker~\clatse{}, which generates an
AAT as an MSC if an attack was found. \tool{} reads the MSC and the
concretization file, and executes the AAT over the real \webapplication{}.
\tool{} makes use of the external tools \wfuzz{}~\cite{wfuzz} and
\sqlmap{}~\cite{sqlmap} in order to perform attacks. 
\end{EXT}
\tool{} is also able
to automatically use information it extracts during an attack (e.g., as a
result of \FileInclusion{}) to proceed further with the execution of an
AAT.
We have run \tool{} on our case studies using a Mac Book laptop (Intel
i5-4288U with 8G RAM and Python3.5). The execution time of the
model-checking phase ranges from 30ms to 50ms, while the overall process
(from MSC generation to concretization) depends on the external tools
\wfuzz{} and \sqlmap{}. 


\subsection{Case study: the \chained{} \webapplication{}}
\label{sec:chained}

\chained{} is a \webapplication{} that we specifically
wrote in order to show how file-system and \sqli{} vulnerabilities can be
combined together to generate multiple attack traces. 
We designed \chained{} to ensure
that it is realistic and representative of software that could indeed be
deployed.

\chained{} implements a typical HTTP login phase in which users can log into the
system by providing username and password. 
The \webapplication{}
performs a query to the database in order to verify the credentials and grant
access to a restricted area. The restricted area allows users to view other
users' profiles and modify their own personal information (Name,
Surname, Phone number),
and it allows for the upload of a personal image to use
as avatar. We check the \webapplication{} for file reading attacks,
i.e., we want to generate multi-stage attack
traces showing how an attacker can exploit file-system and \sqli{}
vulnerabilities to get read access to the 
\webapplication{}'s file-system.
\begin{SHORT}A detailed description of the model can be found in~\cite{wafex}\fix{F}{Extended version}.
\end{SHORT}
\begin{EXT}A detailed description of the model can be found in~\autoref{appendix:casestudy} 
and in~\cite{wafex}.\fix{F}{Mettiamo anche qui riferimento al sito come mi ha
	scritto via email?}
\end{EXT}

We ran this model with \tool{} in order to test the security of \chained{}.
Unfortunately, the model-checker \clatse{} that we use inside \tool{} does not
allow for generating multiple attack traces (nor do the other back-ends of the
AVANTSSAR Platform). Thus, whenever a trace was found, we disabled the branch 
corresponding to the attack
and run \tool{} again to generate another trace different from
the previous one. This process does, of course, miss some traces since by
disabling a branch we prevent any other trace to use that branch in a different
step of the attack trace.\footnote{We plan to extend
	\clatse{} or replace it with a tool capable of generating multiple attack
	traces.} However, it shows that multiple traces can actually be generated.
We generated three different AATs: \#1, \#2 and \#3.

\emph{\textbf{AAT \#1}}
This first AAT shows how an attacker
might be able to exploit an \sqli{} in the login phase to directly read
files from the file-system (\autoref{lst:aat1:chained}). The attacker
\texttt{i} sends to the \webapplication{} a request for login by sending the
payload \texttt{sqli.fsi} (1). The \webapplication{} sends a
query to the database entity (2), which forces the database into sending a
read request to the file-system entity with value \texttt{fsi}
(3). The file-system checks if the provided file is part of the
file-system and answers to the database with that file (4). The database
forwards the response from the file-system
to the \webapplication{} (5), which, finally, sends to the at\-tacker the
\texttt{dashboard} page along with the result from the database (6).


\begin{lstlisting}[breaklines=true,caption={AAT \#1 for accessing the file-system in \chained{}.}, 
		label={lst:aat1:chained}]
i -> WebApp : login.sqli.fsi.Password
WebApp -> DB: query(sqli.fsi)
DB -> FS : readFile(fsi)
FS -> DB : file(fsi)
DB -> WebApp : tuple(file(fsi))
WebApp -> i : dashboard.AuthCookie.tuple(file(fsi))
\end{lstlisting}
%

\emph{\textbf{AAT \#2}}
We disabled the branch that allows
the database to read from the file-system and ran the model again in order
to generate a different AAT (\autoref{lst:aat2:chained}). The attacker
\texttt{i} tries to exploit a \sqli{} in order to write a malicious file
so to exploit a remote code execution.
\texttt{i} sends to the \webapplication{} a request for login by sending the payload \texttt{sqli.evil\_file} (1). The \webapplication{} sends a query to the database entity (2), which forces the database into sending a writing request with value \texttt{evil\_file} to the file-system (3). The
file-system marks the new file as available in the file-system and the
database sends a response to the \webapplication{} with the file just
created (4). The \webapplication{} responds to \texttt{i} with the
\texttt{dashboard} page along with a newly generated cookie and the result of the 
creation of a new file (5). The attacker \texttt{i} now exploits the \texttt{evil\_file} by
sending the payload \texttt{fsi} to the \webapplication{}
(6). The \webapplication{} will now perform a \texttt{readFile()} operation on the file-system and will send the retrieved file back to the attacker (7-9).

\begin{lstlisting}[breaklines=true,caption={AAT \#2 for accessing the file-system in \chained{}.}, 
		label={lst:aat2:chained}]
i -> WebApp : login.sqli.evil_file.Password
WebApp -> DB: query(sqli.evil_file)
DB -> FS : writeFile(evil_file)
DB -> WebApp : tuple(new_file(evil_file))
WebApp -> i : dashboard.AuthCookie.tuple(new_file(evil_file))
i -> WebApp : file.fsi
WebApp -> FS : readFile(fsi)
FS -> WebApp : file(fsi)
WebApp -> i : file(fsi)
\end{lstlisting}


\emph{\textbf{AAT \#3}}
We disabled the branch that allows the database to both read and write from the
file-system, and ran the model again 
to generate a different AAT
(\autoref{lst:aat3:chained}). The attacker \texttt{i} bypasses the
authentication mechanism by sending the \texttt{sqli} payload (1). This allows
him to have access to the \webapplication{}, which responds with a valid
authentication cookie value (2-4). The attacker can now take advantage of the
profile edit page in order to upload a malicious file by sending the \sqli{}
payload \texttt{sqli} and the \texttt{evil\_file} payload (5). The
\webapplication{} sends a query request to the database and the database
answers (6-7). The \webapplication{} now sends a writing request to the
file-system in order to store the newly uploaded avatar \texttt{evil\_file}
(8), and finally the \webapplication{} responds back to the attacker with the
\texttt{profileid} page and the \texttt{tuple(sqli)} resulting from exploiting
a \sqli{} in the editing request (9). The attacker exploits the
\texttt{evil\_file} created in (8),
to read content from the file-system. The \webapplication{} receives a request
for the \texttt{evil\_file} with payload \texttt{fsi} (10) and makes a request
to the file-system for reading \texttt{fsi} (11-12). Finally, the
\webapplication{} sends the file back to the attacker (13).

It is worth remarking what happened in steps (6-7). Since we assumed that all
requests made by the attacker are malicious actions, the only way the attacker
has to proceed is performing a \sqli{} attack in the edit request. However, by
reading the trace it can be easily seen that the \sqli{} is not used to bypass
an authentication or extract information.


\begin{lstlisting}[breaklines=true,caption={AAT \#3 for accessing the file-system in \chained{}.},
		label={lst:aat3:chained}]
i -> WebApp : login.sqli.Password
WebApp -> DB: query(sqli)
DB -> WebApp: tuple(sqli)
WebApp -> i : dashboard.Cookie.tuple(sqli)
i -> WebApp : edit.Name.Surname.sqli.evil_file.Cookie
WebApp -> DB: query(sqli)
DB -> WebApp: tuple(sqli)
WebApp -> FS: writeFile(evil_file)
WebApp -> i : profileid.tuple(sqli)
i -> WebApp : file.fsi
WebApp -> FS: readFile(fsi)
FS -> WebApp: file(fsi)
WebApp -> i : file(fsi)
\end{lstlisting}


\subsection{Concretization phase}
We configured a safe environment where we ran \dvwa{} and our \chained{}
case study. We ran \tool{} and concretized the AATs it generated. The
concretization was successful for all our case studies, 
actually showing how the attacker would perform the real attacks on both \dvwa{} and
\chained{}.
The only example that \tool{} could not concretize is the AAT\#2 in \chained{}.
In that case,
the attacker is supposed to exploit
a \sqli{} for writing to the file-system. \tool{} was not able to
concretize the trace since the user executing the database did not have
the privileges to write to the file-system, which highlights, as we
already stated, that the presence of a vulnerability does not imply
its exploitability and that only a penetration testing phase can analyze
such scenarios.

\begin{SHORT}
\section{Concluding remarks, related work and future work}
\label{sec:concluding}


Our approach is able to find multi-stage attacks to \webapplication{s} that, to
the best of our knowledge, no other tool can find, which involve the
combined exploit of file-system and \sqli{} vulnerabilities.\fix{Luca}{Ho rimesso ``that no other tool can find'' perch\'e senn\`o il resto non seguiva. Lo ho per\`o riscritto con ``to our knowledge''\textbf{F} ok, ho messo ``to the best of our knowledge'' come del resto del paper}
Some related works are, however, worth discussing.
%
%
Pentesting remains the leading methodology for the security analysis of
\webapplication{s}. This is because the human component is crucial in
evaluating the security of the \webapplication{}. Tools like
\wfuzz{}~\cite{wfuzz} or DotDotPwn~\cite{ddp} support the security analyst
in finding the presence of vulnerabilities, but they do not give any clue
on how a vulnerability can be used nor they say if an attack that
uses that vulnerability can actually be carried out.

The idea underlying the methodology for modeling
\webapplication{s} given in~\cite{towards} is similar
to our approach, but they
defined three different attacker models that should find web attacks,
whereas we show how the standard DY attacker can be used. They also
represent a number of HTTP details that we do not require
that eases the
modeling phase. Most importantly, they don't take combination of attacks into consideration.

The model-based security testing tool SPaCiTE~\cite{spacite} starts from a
secure (ASLan++) specification of a \webapplication{} and, by mutating the
specification, automatically introduces security flaws. SPaCiTE implements
a mature concretization phase, but it mainly finds vulnerability entry
points and tries to exploit them, whereas our main goal is to consider how
the exploitation of one or more vulnerabilities can compromise the
security of the \webapplication{}.

The ``Chained Attack'' approach of~\cite{CalviVigano16} considers multiple
attacks to compromise a \webapplication{}, but it does not consider
file-system vulnerabilities nor interactions between vulnerabilities,
which means that it can't reason about using a \sqli{} to access the
file-system. Moreover, it requires an extra effort of the security
analyst, who should provide an instantiation library for the
concretization phase, while we use well-known external state-of-the-art
tools.

The analysis in~\cite{paper:sqli} was limited to \sqli{} for authentication
bypass and data extraction attacks, which we used in this paper as the basis
for considering \sqli{} for accessing the file-system and for modeling the
\chained{} case study.
In~\cite{csrf}, Rocchetto et al. model \webapplication{s} to search for CSRF
attacks. While they limit the analysis to CSRF, 
there could be useful interactions with our approach.

We plan to extend \tool{} (i) with stronger functionalities for the automatic
creation of the \webapplication{} model and (ii) to cover other
complex \webapplication{} vulnerabilities like XSS
and sophisticated multi-stage attacks involving the exploitation of multiple
vulnerabilities.\fix{F}{Ho cambiato l'ultima riga nella quale citavamo homakov 
	perch\'e gi\`a usato da Alberto}
\end{SHORT}

\begin{EXT}
\section{Related work}
\label{sec:related}

To the best of our knowledge, this paper is the first attempt 
to show how model-checking techniques and the standard DY attacker model can be
used for the generation of attack traces where multiple vulnerabilities are
used to violate a security property.  There are, however, previous works that
are closely related to what we presented in this paper and that are thus worth
discussing.


Penetration testing remains the leading methodology for the security
analysis of web applications. This is because the human component is
crucial in evaluating the security of the web application. Tools like
\wfuzz{}~\cite{wfuzz} or DotDotPwn~\cite{ddp} support the security analyst
in finding the presence of vulnerabilities, but they do not give any clue
on how a vulnerability can be used and they do not say if an attack that
uses that vulnerability can actually be carried out.

In~\cite{towards}, Akhawe et al. presented a methodology for modeling web
applications and considered five case studies modeled in the
Alloy~\cite{alloy} language. The idea is similar to our approach, but they
defined three different attacker models that should find web attacks,
whereas we have shown how the standard DY attacker can be used. They also
represent a number of HTTP details that we do not require that eases the 
modeling phave.
Finally, and most importantly, they don't take combination of
attacks into consideration.

In~\cite{spacite}, B\"uchler et al. presented SPaCiTE, a 
model-based security testing tool that starts from a secure (ASLan++)
specification of a web application and, by mutating the specification,
automatically introduces security flaws. SPaCiTE implements a mature
concretization phase, but it mainly finds vulnerability entry points and
tries to exploit them, whereas our main goal is to consider how the
exploitation of one or more vulnerabilities can compromise the security of
the web application.

The ``Chained Attack'' approach of~\cite{CalviVigano16} considered
multiple attacks to compromise a web application. The idea is close to the
one we present in this paper. However, the ``Chained Attack'' approach
does not consider file-system vulnerabilities nor
interactions between vulnerabilities, which means that with that
formalization it would be impossible to represent a \sqli{} to access the
file-system. Finally, the ``Chained Attack'' approach requires an extra
effort of the security analyst, who should provide an instantiation
library for the concretization phase, while we use well-known external
state-of-the-art tools.

The analysis in~\cite{paper:sqli} was limited to \sqli{} for authentication
bypass and data extraction attacks, which we used in this paper as the basis
for considering \sqli{} for accessing the file-system and for modeling the
\chained{} case study.


In~\cite{csrf}, Rocchetto et al. model web applications to search for CSRF
attacks. While they limit the analysis to CSRF, the idea and representation are
close to ours 
and there could be useful interactions between the two approaches.

\section{Conclusions and future work}
\label{sec:conclusionsfuturework}

We have proposed a formalization for the representation of web applications
vulnerable to file-system attacks, and shown how the DY attacker model can be
used in order to find and exploit attacks that violate security properties of
web applications. Our approach is able to find multi-stage attacks to web
applications that, to the best of our knowledge, no other tools can find, which
involve the combined exploit of file-system and \sqli{} vulnerabilities. We
have implemented a prototype tool called \tool{} that takes an \aslanpp{}
specification of a web application and a concretization file, generates an AAT
and automatically tests the resulting AAT against the real web application.
\tool{} does not find payloads for exploiting vulnerabilities but rather makes
use of state-of-the-art tool (\wfuzz{}~\cite{wfuzz} and
\sqlmap{}~\cite{sqlmap}) in order to find the proper payload. As a proof of
concept, we have applied \tool{} to four real-life case studies.

As future work, we plan to extend our approach and \tool{} to cover other
complex web application vulnerabilities such as Cross-Site Scripting as well as
sophisticated multi-stage attacks involving the exploitation of multiple
vulnerabilities.\fix{F}{Ho cambiato l'ultima riga nella quale citavamo homakov 
	perch\'e gi\`a usato da Alberto}
%
%
We also plan to extend
\tool{} by introducing  more functionalities for the automatic creation of the web
application model and the concretization of multi-stage attacks.
\end{EXT}

\bibliographystyle{abbrv}
\bibliography{literature}

\begin{EXT}
\appendix

\section{Database}
\label{appendix:db}
For the sake of completeness, we give here a quick description of the database
formalization we first proposed in~\cite{paper:sqli}.

	The database is defined by an if-elseif statement. In
	line (1), the database accepts messages for a sanitized query expressed by
	the uninterpreted function \texttt{sanitizedQuery()}. When the database
	receives a sanitized query, it is assumed that no \sqli{} is possible and
	thus the query can be performed only if executed with values that are
	stored in the database. The uninterpreted function \texttt{tuple()} is
	used to represent any 
	information 
	stored in the database and (2)
	enforces the constraint that a sanitized query can be performed only on a
	variable of the form \texttt{tuple(*)}, i.e., content stored in the
	database, where \texttt{*} acts as a wildcard that matches any possible
	parameter. The constant \texttt{no\_tuple} is then returned as a response
	from the database to a sanitized query (2). \texttt{no\_tuple} represents
	non-useful knowledge leaking from the database and, 
	since a sanitized query is assumed safe against \sqli{}, it doesn't increase the
	attacker's knowledge.
	
	\begin{lstlisting}[language=pseudo,caption={Pseudo-code representing
		the basic behavior of the database (\texttt{DB}) as given
			in~\cite{paper:sqli}.},label={lst:databaseEntity}]
 if(WebApp -> DB: sanitizedQuery(SQLquery)){
  if(SQLquery == tuple(*)){ DB -> WebApp: no_tuple;
  }
 }elseif(WebApp -> DB: query(SQLquery)){
  if(inDB(SQLquery)){ 
   DB -> WebApp: tuple(SQLquery); 
 }elseif(!(inDB(SQLquery))){
	  DB -> WebApp: no_tuple; 
 }}
	\end{lstlisting}
	
	The main elseif branch (3) defines how the database handles raw queries,
	which are expressed with the uninterpreted function \texttt{query()}. 
	The validity of a query is expressed by the predicate \texttt{inDB()} and the corresponding Horn clause 
	\begin{lstlisting}[language=pseudo,numbers=none]
	db_hc_evil(M): inDB(sqli.M)
	\end{lstlisting}
	where the \sqli{} payload is represented by the constant \texttt{sqli}.
	If the database receives a raw query that is valid, then
	\texttt{tuple(SQLquery)} is sent back (4), else the constant
	\texttt{no\_tuple} is sent back (5).



 \section{The workflow of our approach (and tool)}
 \label{appendix:workflow}
 
 \autoref{fig:workflow} shows the workflow of our approach, which comprises
 five main tasks. \circled{1}~The security analyst uses the Burp proxy to
 record an HTTP trace of requests/responses of the \webapplication{}.
 \circled{2} The security analyst uses our model generator in order to
 translate the HTTP trace recorded by Burp into an \aslanpp{} model.
 \circled{3} The security analyst completes the generated model in order to
 define the behavior and security goals he wants to test. \circled{4}
 \tool{} takes in input the \aslanpp{} model that the security analyst has
 generated and invokes \clatse{}~\cite{atse}, which model-checks the model
 and searches for an attack trace that violates the security goal defined
 by the security analyst. \clatse{} generates an AAT as an MSC if an attack
 trace was found, which is then used by \tool{}, along with the
 concretization file, as input for the concretization engine that uses
 state-of-the-art-tools such as \sqlmap{} and \wfuzz{} to concretize the
 AAT and apply it to test the real \webapplication{} \circled{5}.
 
 \begin{figure}[t]
 \begin{center}
 \includegraphics[width=\textwidth]{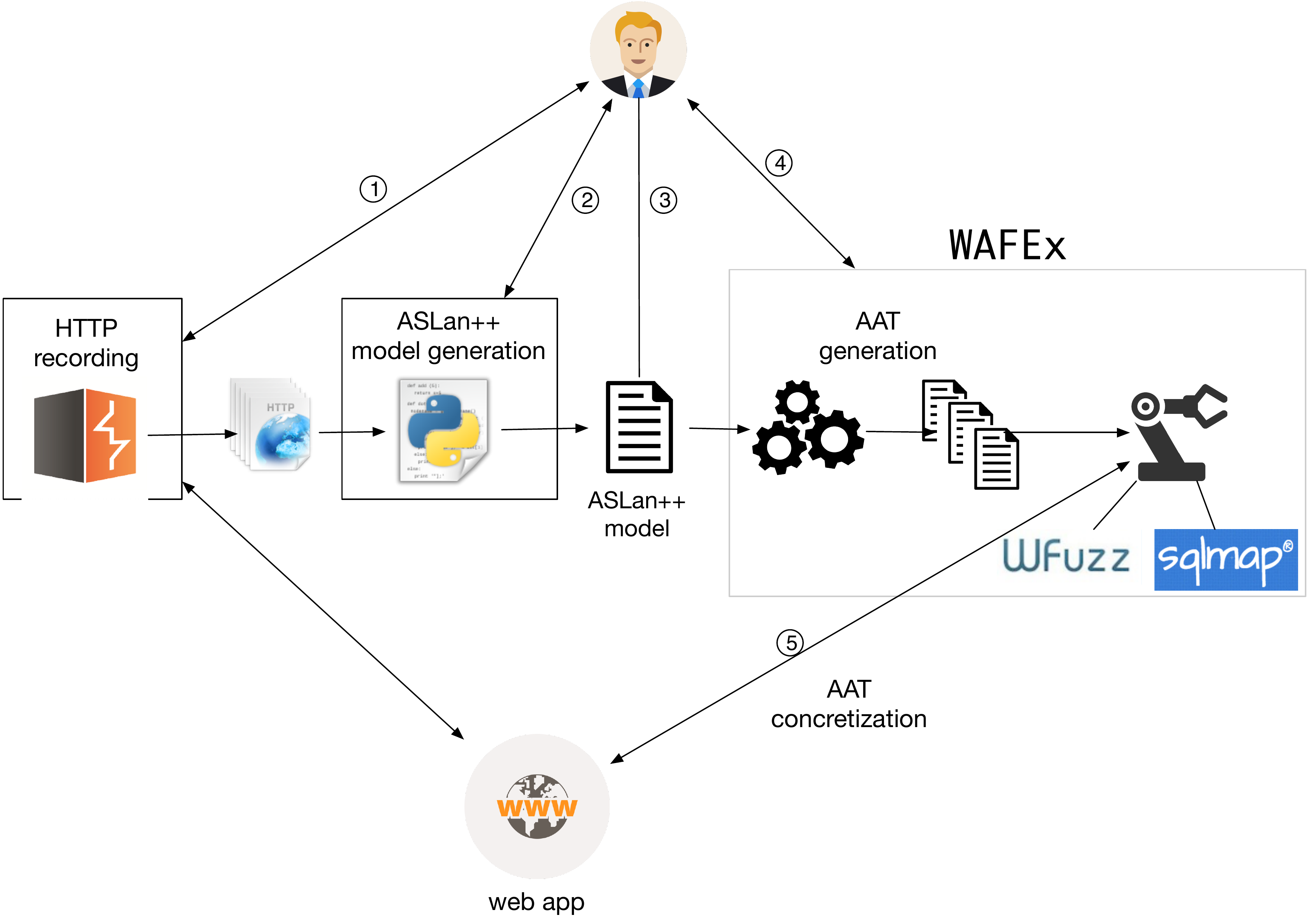}
 \caption{Workflow of our approach (and tool)}
 \label{fig:workflow}
 \end{center}
 \end{figure}

 \section{Case studies}
 \label{appendix:casestudies}
 
 
 \subsection{File Inclusion (\dvwa{})}
 
 \emph{Damn Vulnerable Web Application (\dvwa{})}~\cite{dvwa} is a
 vulnerable \webapplication{} that provides an environment in which
 security analysts can test their skills and tools. \dvwa{} is divided in
 examples implementing web pages vulnerable to the most common
 \webapplication{} vulnerabilities. For the first case study, we consider
 the \FileInclusion{} scenario, which is an example of a \webapplication{}
 vulnerable to \FileInclusion{} just like the running example we used
 in~\autoref{sec:formalization}. The model is the one given
 in~\autoref{lst:includeExample} and as goal we used the one
 in~\autoref{lst:goals}.\footnote{In our case studies, we use the variable
 \texttt{Client} when the \webapplication{} is expecting an interaction
 with a client. Since we assumed only malicious interactions with the
 \webapplication{} (\autoref{sec:formalization}), the variable
 \texttt{Client} will always be instantiated with the DY attacker
 \texttt{i} in concrete \webapplication{} executions.}
 
 \smallskip\emph{\textbf{Abstract Attack Trace}}
 
 We ran the model with \tool{}, which generated the AAT in
 \autoref{lst:aat:fileinclusion}, which shows an attack that exploits a
 \FileInclusion{} vulnerability allowing the attacker \texttt{i} to get
 access to the file-system. The attacker \texttt{i} sends a request for the
 include page along with a file inclusion payload represented by the
 constant \texttt{fsi} (1). The \webapplication{} performs a read request
 to the file-system (2), which responds to the \webapplication{} with the
 requested file (3), and the \webapplication{} sends back the file (4) to
 the attacker.

 \begin{lstlisting}[caption={AAT that exploits a \FileInclusion{} vulnerability.}, label={lst:aat:fileinclusion}]
 i -> WebApp : include.fsi
 WebApp -> FS: readFile(fsi)
 FS -> WebApp: file(fsi)
 WebApp -> i : include.file(fsi)
 \end{lstlisting}
 
 \subsection{Unrestricted File Upload (\dvwa{})}
 
 In this case study, we used the \UnrestrictedFileUpload{} example from
 \dvwa{} to show how simple it can be to represent an \UnrestrictedFileUpload{} vulnerability with our formalization.
 The example is indeed quite straightforward: a \webapplication{} allows
 users to upload files. The model is shown in \autoref{lst:fileupload}. 
 When it receives a request for file upload (1), the \webapplication{}
 makes a request to the file-system entity for writing the file (2)
 and, since we assumed a writing operation always succeeds, the
 \webapplication{} answers back to the client with the \texttt{uploaded}
 page (3). As goal, in \autoref{lst:goal:fileupload} we check that a
 malicious file, represented by the constant \texttt{evil\_file}, is not
 part of the file-system.
 
 \smallskip\emph{\textbf{Abstract Attack Trace}}
 We ran the model with \tool{}, which generated the AAT in
 \autoref{lst:aat:fileupload}, which shows an attack that exploits an
 \UnrestrictedFileUpload{} vulnerability. The attacker \texttt{i} sends a
 request for uploading a file by sending the \texttt{evil\_file} payload
 representing a malicious server side code (1). The \webapplication{} sends
 a writing request to the file-system (2) and answers back to the client
 with the \texttt{uploaded} page. At the end of this execution, the file
 represented by the constant \texttt{evil\_file} will be part of the
 file-system.
 
 \begin{lstlisting}[caption={Pseudo-code representing an \UnrestrictedFileUpload{} vulnerability.},
 		label={lst:fileupload}]
 if(Client -> WebApp : uploaded.File){
  WebApp -> FS : writeFile(File);
  WebApp -> Client : uploaded;}
 \end{lstlisting}
 
 \begin{lstlisting}[numbers=none,caption={Goal representing that a malicious file has been uploaded to the web 
 				application's file-system.},
 		label={lst:goal:fileupload}]
 [](!inFS(evil_file));
 \end{lstlisting}
 
 \begin{lstlisting}[caption={AAT that exploits an \UnrestrictedFileUpload{} vulnerability.},
 		label={lst:aat:fileupload}]
 i -> WebApp : uploaded.evil_file
 WebApp -> FS: writeFile(evil_file)
 WebApp -> i : uploaded
 \end{lstlisting}

 \subsection{\sqli{} read (\dvwa{})}
 
 In this case study, we used the \sqli{} example from \dvwa{} to show how
 it is possible to exploit a \sqli{} for reading content from the
 file-system. This example from DVWA was meant to show how a \sqli{}
 vulnerability can be exploited for extracting data from the database.
 However, \tool{} was able to generate a wider set of attack traces showing
 that the \sqli{} in DVWA could also be used for reading files from the
 file-system. Consider a \webapplication{} that accepts a value
 representing a user id that is used in a \texttt{SELECT} query for
 retrieving and showing details about that user. The model is given in
 \autoref{lst:sqliread}. The \webapplication{} accepts a request for the
 \texttt{userid} page along with the \texttt{IDvalue} value (1). The
 \webapplication{} then performs a query to the database by sending a query
 request (2). The \webapplication{} waits to receive back the result of the query
 from the database (3) and, upon receiving \texttt{tuple(Response)}, it
 sends to the user the page \texttt{userid} and the result of the query
 \texttt{tuple(Response)}. \autoref{lst:goal:sqliread} shows the 
 file-system access goal that we defined in which the attacker should not
 be able to know \texttt{file()}.
 
 \begin{lstlisting}[caption={Pseudo-code representing a \sqli{} for file reading vulnerability.},
 		label={lst:sqliread}]
 if(Client -> WebApp : userid.IDvalue){
  WebApp -> DB : query(IDvalue);
  if(DB -> WebApp : tuple(Response)){
   WebApp -> Client: userid.tuple(Response);}}
 \end{lstlisting}
 
 \begin{lstlisting}[numbers=none,caption={Goal representing that the attacker knows the content of the file-system.},
 		label={lst:goal:sqliread}]
 [](!iknowledge(file(*)));
 \end{lstlisting}
 
 \smallskip\emph{\textbf{Abstract Attack Trace}}
 We ran the model with \tool{}, which generated the AAT in
 \autoref{lst:aat:sqliread}, which shows an attack in which the attacker
 \texttt{i} is able to exploit a \sqli{} vulnerability in order to read from the file-system. The attacker \texttt{i} sends the concatenation of \texttt{userid} with the \sqli{} payload \texttt{sqli} and the \texttt{fsi} payload (1). The \webapplication{}
 sends a query to the database (2), which, thanks to our formalization, 
 can be forced into performing a reading operation to the file-system (3).
 The file-system checks that the file is a valid file and answers back to
 the database wrapping the \texttt{fsi} payload with the \texttt{file()}
 function (4). The database wraps the answer from the file-system with
 \texttt{tuple()} and sends it back to the \webapplication{} (5). Finally,
 the \webapplication{} forwards back to the attacker the result of the
 query along with the \texttt{userid} page (6).
 
 \begin{lstlisting}[breaklines=true,caption={AAT that exploits a \sqli{} for reading files from the file-system.}, label={lst:aat:sqliread}]
 i -> WebApp : userid.sqli.fsi
 WebApp -> DB : query(sqli.fsi)
 DB -> FS : readFile(fsi)
 FS -> DB : file(fsi)
 DB -> WebApp : tuple(file(fsi))
 WebApp -> i : userid.tuple(file(fsi))
 \end{lstlisting}

 \subsection{\chained{}}
\label{appendix:casestudy} 
\chained{} implements a typical HTTP login phase in which users can log into the
system by providing username and password credentials. The \webapplication{}
performs a query to the database in order to verify the credentials and grant
access to a restricted area. The restricted area allows users to view other
users' profiles and modify their own personal information, which are Name,
Surname, Phone number, and it allows for the upload of a personal image to use
as avatar. We check the \webapplication{} for file reading attacks
i.e., we want to generate multi-stage attack
traces showing how an attacker can exploit file-system and \sqli{}
vulnerabilities 
\webapplication{}'s file-system.

\autoref{lst:chained} shows a model for such a \webapplication{}, which
has three branches of an if-elseif statement, each one responding to a
different request.

\begin{lstlisting}[breaklines=true,caption={Pseudo-code representing the \chained{} \webapplication{}.},
		label={lst:chained}]
if(Client -> WebApp : login.Username.Password){
 WebApp -> DB : query(Username.Password);
 if(DB -> WebApp : tuple(SQLresponse){
  AuthCookie := fresh();
  sessionValue(AuthCookie);
  WebApp -> Client : dashboard.tuple(SQLresponse).AuthCookie;}
}elseif(Client -> WebApp : profileid.Id.AuthCookie){
 if(sessionValue(AuthCookie){
  WebApp -> DB: query(Id);
  if(DB -> WebApp: tuple(SQLresponse)){
   WebApp -> Client: profileid.tuple(SQLresponse);}}
}elseif(Client -> WebApp : edit.Name.Surname.Phone.Avatar.AuthCookie){
 if(sessionValue(AuthCookie)){
 WebApp -> DB : query(Name.Surname.Phone.Avatar);
 if(DB -> WebApp : tuple(SQLresponse)){
  WebApp -> FS : write(Avatar);
  WebApp -> Client : profileid.tuple(SQLresponse);}}
 }elseif(Client -> WebApp : evil_file.FilePath){
  if(inFS(evil_file){
   WebApp -> FS : readFile(FilePath);
   if(FS -> WebApp : file(FilePath)){ WebApp -> Client: file(FilePath);}}}
\end{lstlisting}

The first branch (1) represents how the \webapplication{} responds to a
login request in which the client sends credentials (\texttt{Username} and
\texttt{Password}). When the \webapplication{} receives the login request,
it performs a query to the database in order to verify the received
credentials (2). If the credentials are correct, then the database
responds back with \texttt{SQLresponse} wrapped with the function
\texttt{tuple()} (3), and then the \webapplication{} creates a variable
\texttt{AuthCookie} (4), marks it as a session value by using the
\texttt{sessionValue} predicate (5), and finally responds to the client by
sending the \texttt{dashboard} page, the result of the query
\texttt{tuple(SQLresponse)} and the cookie \texttt{AuthCookie} (6).

The second branch (7) responds to the request to view a user's profile.
The client sends a request to the \webapplication{} for the
\texttt{profileid} page with the \texttt{Id} value for the user's profile
to view and a \texttt{AuthCookie} value that is used to verify that the
request comes from an authenticated user. The \webapplication{} first
checks if the cookie provided in the request is a valid session value (8)
and, if it is, then the \webapplication{} performs a query to the database
by sending the \texttt{Id} of the profile that has been requested (9). If
the database sends the result of the executed query
\texttt{tuple(SQLresponse)} (10), then the \webapplication{} responds to
the client by sending the \texttt{profileid} page and
\texttt{tuple(SQLresponse)} (11).

The third branch handles the edit profile request. The client sends a
request to the \webapplication{} for the editing page \texttt{edit} along
with values for \texttt{Name}, \texttt{Surname}, \texttt{Phone},
\texttt{Avatar} and an \texttt{AuthCookie} value (12). The
\webapplication{} checks if the cookie provided in the request is a valid
session value (13), and if it is, then the \webapplication{} performs a
request to the database to edit the values for \texttt{Name},
\texttt{Surname}, \texttt{Phone} and \texttt{Avatar} (14). Since editing
profile values is performed with an \texttt{UPDATE} query, the \webapplication{} checks that the answer from the database is
\texttt{tuple(SQLresponse)} (15), and then performs a writing operation to
the file-system to save the \texttt{Avatar} value (16). Since we assumed
the writing operation to always succeed, the \webapplication{} sends,
without further checks, the page \texttt{profileid} and the result of the
query \texttt{tuple(SQLresponse)} to the client (17).

The last part of the model shows how to deal with remote code execution
(\autoref{sec:webapp}). We include in the specification of the
\webapplication{} a typical server-side code that would allow an attacker
to have reading access over the \webapplication{}'s file-system. The
\webapplication{} can receive a request for \texttt{evil\_file} (18) only
if \texttt{evil\_file} is part of the file-system (19). If that is the
case, then \texttt{evil\_file} allows an attacker to perform a reading
request to the file-system (20), where the \texttt{FilePath} variable is
used to specify the file that has to be retrieved. If the file-system
responds back to the \webapplication{} with \texttt{file(FilePath)} (21),
then the \webapplication{} will forward \texttt{file(FilePath)} back to
the client (22).

We check the \webapplication{} for file reading attacks
(\autoref{lst:goal:chained}).

\begin{lstlisting}[numbers=none,caption={Goal representing that the attacker knows the content of the file-system.},
		label={lst:goal:chained}]
[](!(iknowledge(file(*))));
\end{lstlisting}

\smallskip\emph{\textbf{Abstract Attack Trace}}
See~\autoref{sec:chained}.
 
\section{Implementation in ASLan++}
\label{appendix:implementationaslanpp}

We now present the details of how we implemented our formalization using
the formal language ASLan++. It is worth remembering that our approach is
not strictly related to ASLan++, in fact our approach is general enough
that it could be quite straightforwardly used with other specification
languages and/or other reasoners implementing the Dolev-Yao attacker model.

\subsection{Skeleton model}
 \label{sec:skeleton}
 We now present the ASLan++ code of a skeleton model that implements all
 the aspects described in \autoref{sec:formalization}. This skeleton is
 intended to be a base ASLan++ model from which a security analyst can
 start the creation of a \webapplication{} model. We first describe agents,
 variables, constants, facts and uninterpreted functions
 (\autoref{lst:symbols}), and then describe the clauses that define the
 behavior of \sqli{} and file-system attacks (\autoref{lst:clauses}), the
 file-system entity (\autoref{lst:db}), the database entity
 (\autoref{lst:fs}) and the \webapplication{} entity (\autoref{lst:webApp}).
 
 We assume that the reader is familiar with the syntax of ASLan++; details
 can be found, for instance, in~\cite{avantssar-tacas,spacios} and in the documents referenced there.
 
 \begin{lstlisting}[breaklines=true, caption={ASLan++ code of the symbols used in the skeleton of a \webapplication{}.},label={lst:symbols}]
 specification Skeleton
 channel_model CCM
 entity Environment{
 symbols
 ▸webapplication,database,filesystem:agent;
 
 % Malicious payload
 ▸sqli,fsi,evil_file:text;
 
 % Database
 ▸nonpublic inDB(message):fact;
 ▸nonpublic query(message):message;
 ▸nonpublic sanitized(message):message;
 ▸nonpublic tuple(message):message;
 ▸nonpublic no_tuple:text;
 
 % Filesystem
 ▸nonpublic readFile(message):message;
 ▸nonpublic file(message):message;
 ▸nonpublic inFS(message):fact;
 ▸nonpublic isInFS(message):fact;
 ▸nonpublic writeFile(message):message;
 ▸nonpublic no_file:text;
 
 % Sessions
 ▸nonpublic sessionValue(message):fact;
 
 % request¬
 ▸http_request(message,message,message):message;
 % response
 ▸http_response(message,message):message;
 % separator
  s, none : text;
 \end{lstlisting}
 
 Lines 1--3 begin the skeleton specification by stating a name
 for the specification (\texttt{Skeleton}), the channel model used (\texttt{CCM})
 and by introducing the outermost entity (\texttt{Environment}).
 The symbols section of the environment entity begins in line 4. We define
 the constants representing the agents involved in the communication (5): the \webapplication{} (\texttt{webapplication}), the database (\texttt{database}) and the file-system (\texttt{filesystem}).
 In line 8, we define the abstract payload that the DY attacker will use
 for attacking the \webapplication{}: \texttt{sqli} for \sqli{} attacks,
 \texttt{fsi} for file-system attacks and \texttt{evil\_file} for executing arbitrary server-side code.
 In lines 11--15, we define predicates and uninterpreted functions used for
 implementing the database behavior as described in~\autoref{sec:db}.
 In lines 18--23, we define predicates and uninterpreted functions used for
 implementing the file-system behavior as described in~\autoref{sec:fs}.
 In line 26, we define the predicate that is used to describe when a constant is a session value.
 In lines 29 and 31, we define two uninterpreted functions
 (\texttt{http\_request} and \texttt{http\_response}) that are used to define an HTTP request and an HTTP response, respectively. 
 In line 32, we define the constant \texttt{s}, which is used as separator in the list of parameters of an HTTP request, and the constant \texttt{none}, which is used for optional parameters.
 
 The uninterpreted function \texttt{http\_request} has three parameters:
 (1) the web page being requested, (2) the list of parameters and (3) a
 cookie value. The web page is defined with a constant, while the list of
 parameters as a concatenation of text of the form \texttt{const.s.Val},
 where \texttt{const} is a constant representing a key and \texttt{Val} is
 a variable representing the value for the \texttt{const} key (and
 \texttt{s} is used as separator). The cookie value is optional and is
 represented with a constant.
 
 The uninterpreted function \texttt{http\_reponse} has two parameters: (1)
 the web page sent back to the client and (2) the content of the response,
 which is a concatenation of constants, variables and uninterpreted
 functions that represent information that needs to be sent in the
 response, e.g., a cookie value or the function \texttt{file()}.
 
 Our skeleton defines the behavior of \sqli{} and file-system
 attacks, as described in \autoref{sec:db} and \autoref{sec:fs},
 with a series of Horn clauses, shown in \autoref{lst:clauses}. 
 
 \begin{lstlisting}[breaklines=true, caption={ASLan++ code of the 
 Horn clauses used in the skeleton.},label={lst:clauses}]
 %DATABASE (behavior)
 db_hc_ev(M) : inDB((sqli.?).M);
 db_hc_ev_2(M) : inDB(sqli.M);
 
 %FILESYSTEM (behavior)
 fs_hc_ev(M) : inFS(fsi.M);
 fs_hc_ev_2(M) : inFS((fsi.?).M);
 fs_hc(M) : inFS(M) :- isInFS(M);
 \end{lstlisting}
 
 Lines 2--3 define clauses for \sqli{} attacks, 6--8 define the
 behavior of a file-system attack. It is worth noticing two things.
 First, in ASLan++, if we write \texttt{sqli.M} we mean a message that is a
 composition of exactly 2 texts. Lines 3 and 7 are used to define an arbitrary arity for a message payload.
 Second, in order to have the Horn clause to evaluate to true whenever a
 new file is added to the file-system, in line 8, we define that
 \texttt{inFS()} is true for a message \texttt{M} whenever the predicate
 \texttt{IsInFS()} for the message \texttt{M} is true. We use the
 \texttt{IsInFS()} predicate in the writing branch of the file-system
 entity in order to state that a new file is now part of the file-system
 (\autoref{lst:fs}).
 
 Our skeleton defines three entities: the file-system entity
 (\autoref{lst:fs}), the database entity (\autoref{lst:db}) and the \webapplication{} entity (\autoref{lst:webApp}). These three entities are
 subentities of the session entity that in ASLan++ is generally used for
 instantiating the model.
 
 The \texttt{Filesystem} entity in \autoref{lst:fs} follows the pseudo-code
 that we presented in \autoref{sec:fs}. The while loop in line 8 wraps the
 entity body content and defines that the file-system entity is actively
 listening for incoming communications. Upon receiving a reading operation
 (10), if the predicate \texttt{inFS()} holds for the variable
 \texttt{Path} (11), the file-system answers back with \texttt{file(Path)}
 (12). If the predicate \texttt{inFS()} doesn't hold (14), the constant
 \texttt{no\_file} is sent back instead (15). Two things are worth
 noticing: the use of \texttt{select-on} and the introduction of nonces. We
 implemented the two main branches as \texttt{select-on} since in ASLan++
 the semantics of \texttt{select-on} saves one or more transitions with
 respect to \texttt{if} when the ASLan++ specification is translated into a
 transition system. The introduction of nonces in any on-going message is
 used to avoid spurious man-in-the-middle attacks between the entities.
 
 \begin{lstlisting}[breaklines=true, caption={ASLan++ code of the file-system	entity.},label={lst:fs}]
 entity Session(Webapplication, Database, Filesystem: agent) {
 entity Filesystem(Webapplication, Actor: agent){
 symbols
  Nonce: text;
  Path : message;
  Entity : agent;
 body{
 while(true){
  select{
   on(?Entity *->* Actor : readFile(?Path).?Nonce):{
    select{on(inFS(Path)):{
     Actor *->* Entity : file(Path).Nonce;
 }
   on(!inFS(Path)):{
    Actor *->* Entity : no_file.Nonce;
 }}}
   on(?Entity *->* Actor : writeFile(?Path).?Nonce):{
    isInFS(Path);
 }}}}}
 \end{lstlisting}
 
 The \texttt{Database} entity in \autoref{lst:db} implements the behavior
 described in \autoref{sec:db}. We introduce two more branches that define
 the communication between the database and the file-system entity. Lines
 20--26 define that the database can communicate with the file-system
 entity to perform a reading operation. Upon receiving a SQL query, in case
 the Horn clause \texttt{inDB()} holds (20), the database can perform a
 reading operation on the file-system (22-23) and, if the file-system
 answers back with a file (24), the database entity will wrap the answer
 with the uninterpreted function \texttt{tuple()} and forward it back to
 the \webapplication{} (25). Lines 27--32 define that the database can
 communicate with the file-system entity to perform a writing
 operation. Upon receiving a SQL query, in case the Horn clause
 \texttt{inDB()} holds (27), the database can perform a writing operation
 on the file-system (28-31). It is worth noticing the introduction of the
 uninterpreted function \texttt{newFile()}. When the database is performing
 a writing operation, it will send back to the \webapplication{} the
 uninterpreted function \texttt{tuple()} wrapping the uninterpreted
 function \texttt{newFile()}, which notifies the \webapplication{} the
 creation of a new file.
 
 \begin{lstlisting}[breaklines=true, caption={ASLan++ code of the database
 				entity.},label={lst:db}]
 entity Database(Webapplication,Actor,Filesystem: agent){
 symbols
  NonceWA,NonceDB,NonceFS: text;
  SQLquery, File: message;
  Sql : message;
 body{
 while(true){
 select{
  on(Webapplication *->* Actor:
   query(sanitized(?SQLquery)).?NonceWA):{
    select{on(SQLquery = tuple(?) 
      & SQLquery != tuple(file(?))):{
       Actor *->* Webapplication:tuple(?).NonceWA;
  }}}
  on(Webapplication *->* Actor:query(?SQLquery).?NonceWA):{
   select{
   on(inDB(SQLquery)):{
    Actor *->* Webapplication:tuple(SQLquery).NonceWA;
   }
  on(inDB(SQLquery)):{
   NonceDB := fresh();
   select{on(SQLquery=(?Sql.?File).?):{
    Actor *->* Filesystem: readFile(File).NonceDB;
    select{on(Filesystem *->* Actor:file(File).NonceDB):{
    Actor *->* Webapplication:tuple(file(File)).NonceWA;
  }}}}}
  on(inDB(SQLquery)):{
   NonceDB := fresh();
   select{on( SQLquery=(?Sql.?File).?):{
    Actor *->* Filesystem:writeFile(File).NonceDB;
    Actor *->* Webapplication:tuple(newFile(File)).NonceWA
  }}}
  on((!inDB(SQLquery) & 
   SQLquery != sanitized(?)) ):{
    Actor *->* Webapplication:no_tuple.NonceWA;
 }}}}}}}
 \end{lstlisting}

 The \texttt{Webapplication} entity in \autoref{lst:webApp} is written as a
 series of \texttt{select-on} branches, where the guard of each
 \texttt{select-on} defines an HTTP request the \webapplication{} can
 handle and the body for that branch defines the behavior of the
 \webapplication{} upon receiving that HTTP request. This is the only part
 of our skeleton that changes accordingly to the \webapplication{} being
 analyzed.

 \begin{lstlisting}[breaklines=true, caption={ASLan++ code of the web application	entity.},label={lst:webApp}]
 entity Webapplication(Actor,Database,Filesystem: agent){
 symbols
  %all the symbols used in the body of this
  %entity in the body below
 body{ %write the behavior of the web app.
  %select{
  % on( ? -> Actor):{
  %  do something here;
  % } ...
 }}}
 \end{lstlisting}
 
 The final part of the skeleton (\autoref{lst:session}) implements the
 instantiation of the \webapplication{}, file-system and database entities
 carried out by the session entity (1-5), the definition of the goals (6-8)
 and the instantiation of the session entity carried out by the environment
 entity (9-11).
 
 \begin{lstlisting}[breaklines=true, caption={ASLan++ code of the session body.},label={lst:session}]
 body{
  new Webapplication(webapplication,database,filesystem);
  new Database(webapplication,database,filesystem);
  new Filesystem(webapplication,filesystem);
 }
 goals %of session¬
  goal_label: % put your security goal here
 }
 body{ %of Environment
  new Session(webapplication, database, filesystem);
 }}
 \end{lstlisting}
 
 \subsection{ASLan++ specifications of our case studies}
 \label{sec:aslanppcasestudies}
 
 In this section, we give the ASLan++ codes that implement the DVWA case studies
 and the \chained{} case study we presented in \autoref{sec:casestudies}. The
 ASlan++ codes of this section are meant to fill the empty spaces of the ASLan++
 skeleton for the \webapplication{} entity in \autoref{sec:skeleton}.

 \subsubsection{File inclusion (DVWA)}
 We used the lesson named ``File Inclusion'' from Damn Vulnerable Web
 Application (DVWA)~\cite{dvwa}. The \webapplication{} entity
 (\autoref{lst:fileInclusion}) has one branch that defines the behavior of
 the \webapplication{} upon receiving a request for including a file. Lines
 1--3 define symbols used by the \webapplication{}. Lines 6--8 define that
 the \webapplication{} can handle a request for page \texttt{include} with
 parameters \texttt{page.s.?Path} and without cookie
 (\texttt{none}).\footnote{The constant \texttt{tag1} is used for
	 concretization purposes, see \autoref{appendix:concretization} for further
 details.} In line 9, a nonce is generated to ensure a fresh communication,
 in lines 10--11, the \webapplication{} performs a reading action on the
 file-system and in line 12, the \webapplication{} sends an HTTP response
 back to the client with the \texttt{include} page and \texttt{file(Path)}.

 \begin{lstlisting}[breaklines=true, caption={ASLan++ code for the ``File Inclusion''	lesson of DVWA.},label={lst:fileInclusion}]
 symbols
  IP: agent;
  NonceWA,NonceDB:text;
 body{
 while(true){
 select{
 % implementing include functionality
  on( ? *->* Actor : ?IP.http_request(include,page.s.?Path, none).tag1):{
   NonceWA := fresh();
   Actor *->* Filesystem : readFile(Path).NonceWA;
   Filesystem *->* Actor : file(Path).NonceWA;
   Actor *->* IP : http_response(include,file(Path));
 }}}}
 \end{lstlisting}
 
 As goal (\autoref{lst:goal:fileInclusion}), we define that the attacker
 should not be able to access something that is function of \texttt{file()}.
 
 \begin{lstlisting}[breaklines=true, caption={ASLan++ code for the file-system access of the ``File Inclusion'' lesson of DVWA.},label={lst:goal:fileInclusion}]
 [](!(iknows(file(?))));
 \end{lstlisting}
 
 \subsubsection{Upload (DVWA)}
 We used the lesson named ``Upload'' from DVWA. The \webapplication{}
 entity (\autoref{lst:upload:real}) has one branch that defines the
 behavior of the \webapplication{} upon receiving a request for uploading a
 file. Lines 1--3 define symbols used by the \webapplication{}. Lines 6--10
 define that the \webapplication{} can handle a request for page
 \texttt{upload} with parameters \texttt{file.s.?Path} and without cookie
 (\texttt{none}). In line 8, a nonce is created to ensure a fresh
 communication and in line 9, the \webapplication{} communicates with the
 file-system for writing the file \texttt{Path} to the file-system.
 Finally, the \webapplication{} sends back to the client an HTTP response
 with the page \texttt{upload} and the uploaded file \texttt{file(Path)}.
 It is worth noticing that the \webapplication{} does not have to wait for
 an answer from the file-system for the writing operation since we assumed
 that no access policies are in place and thus any writing operation will
 always succeed (see~\autoref{sec:fs}).
 
 \begin{lstlisting}[breaklines=true, caption={ASLan++ code for the ``Upload'' lesson of DVWA.},label={lst:upload:real}]
 symbols
 IP: agent;
 NonceWA:text;
 body{
 while(true){
 select{
  on( ? *->* Actor:?IP.http_request(upload,file.s.?Path,none).tag1):{
   NonceWA := fresh();
   Actor *->* Filesystem:writeFile(Path).NonceWA;
   Actor *->* IP:http_response(upload,file(Path));
 }}}}
 \end{lstlisting}
 
 As goal (\autoref{lst:upload}) we check that a malicious file, represented by the constant \texttt{evil\_file} is not part of the file-system.
 \begin{lstlisting}[breaklines=true, caption={ASLan++ code for the file-system attack of the ``Upload'' lesson of DVWA.},label={lst:upload}]
 [](!(inFS(evil_file)));
 \end{lstlisting}
 
 \subsubsection{\sqli{} read (DVWA)}
 We used the lesson named ``SQL Injection'' from DVWA. The
 \webapplication{} entity (\autoref{lst:sqli}) has one branch that defines
 the behavior of the \webapplication{} upon receiving a request for
 querying the database. Lines 1--4 define symbols used by the
 \webapplication{}. Lines 7--9 define that the \webapplication{} can handle
 a request for page \texttt{viewUser} with parameters
 \texttt{userId.s.?IDvalue} and without a cookie (\texttt{none}). In Line
 9, a nonce is generated for ensuring a fresh communication, in lines
 10--11, the \webapplication{} performs a SQL query communicating with the
 database entity and in lines 12--13, the \webapplication{} verifies that
 \texttt{tuple()} is sent back from the database and sends an HTTP response
 back to the client.
 
 \begin{lstlisting}[breaklines=true, caption={ASLan++ code for the file-system access of the ``File Inclusion'' lesson of DVWA.},label={lst:sqli}]
 symbols
 IDvalue, SQLquery, SQLresponse: message;
 IP: agent;
 NonceWA : text;
 body{
 while(true){
 select{
  on(? *->* Actor: ?IP.http_request(viewUser,userId.s.?IDvalue, none).tag1 ):{
   NonceWA := fresh();
   SQLquery := IDvalue;
   Actor *->* Database : query(SQLquery).NonceWA;
   select{on(Database *->* Actor : tuple(?SQLresponse).NonceWA):{
   Actor *->* IP: http_response(viewUser, tuple(SQLresponse));
 }}}}}}
 \end{lstlisting}
 
 As goal we use the same goal that we defined
 in~\autoref{lst:goal:fileInclusion} that states the attacker should not be
 able to access something that is function of \texttt{file()}.
 
 \subsubsection{\chained{}} 
 The ASLan++ model for the \webapplication{} described in
 \autoref{sec:chained} is given in \autoref{lst:webappchained}. Lines 1-6,
 define symbols used by the \webapplication{}. The body of the
 \webapplication{} entity (lines 7--47) describes how the \webapplication{}
 can handle four different HTTP requests.
 
 The first branch (10--19) handles a login process, where the
 \webapplication{} receives a request for page \texttt{index} with
 parameters \texttt{usr.s.?Username.s.pwd.s.?Password} and without cookie
 (\texttt{none}). The \webapplication{} then creates a nonce to ensure a
 fresh communication (11) and queries the database with the provided
 username and password (12--13). If the \webapplication{} receives
 \texttt{tuple(?SQLresponse)} as a response for the query (15), it
 instantiates a variable \texttt{AuthCookie} to a fresh value, marks it as
 a session value and sends it back to the client in an HTTP response
 (16--18).
 
 The second branch (20--27) handles the possibility for a logged-in user to
 view another user's profile, where the \webapplication{} receives a
 request for page \texttt{profileId} with parameters \texttt{id.s.?Id} and
 cookie value \texttt{?AuthCookie}. The \webapplication{} checks if the
 value of \texttt{AuthCookie} is actually a valid session value (21) and,
 if that is the case, performs a query to the database (22--24). The
 \webapplication{} verifies to receive \texttt{tuple(SQLresponse)} as
 answer from the query (25) and, if that is the case, sends an HTTP
 response back to the client with the \texttt{profileId} page, the result
 of the executed query (\texttt{tuple(SQLresponse)}) and the session cookie
 value \texttt{AuthCookie} (26).
 
 The third branch (28--37) handles the possibility for a logged-in user to
 edit a profile where the \webapplication{} receives a request for page
 \texttt{edit} with parameters
 \texttt{name.s.?Name.s.surname.s.?Surname.s.phone.s.?Phone.s.avatar.s.?Avat
 ar} and cookie value \texttt{?AuthCookie}. The \webapplication{} checks if
 the value of \texttt{AuthCookie} is actually a valid session value (29)
 and, if that is the case, performs a query to the database for editing the
 user's details (30--32). If the database answers with
 \texttt{tuple(?SQLresponse)} (33), the \webapplication{} communicates with
 the file-system in order to save the avatar file just uploaded (34-35),
 and finally answers to the client with an HTTP response redirecting the
 user to the page \texttt{profileId} and providing the result of the
 executed query and the session cookie value (36).
 
 The fourth branch (40--47) simulates the presence of a malicious server
 side script. Line 40 defines that the \webapplication{} can answer to a
 request for page \texttt{evil\_file} with parameters \texttt{file.s.?Path}
 and no cookie (\texttt{none}) and line 41 specifies that the page
 \texttt{evil\_file} should be part of the file-system and the variable
 \texttt{?Path} is different from \texttt{evil\_file} (this aims to avoid
 spurious traces, where the DY attacker sends requests of the form
 \texttt{http\_request(evil\_file,file.s.evil\_file,none)}, where he
 exploits \texttt{evil\_file} for reading \texttt{evil\_file}).
 
 If the conditions on line 41 are satisfied, it means that
 \texttt{evil\_file} has been uploaded on the remote file-system and that
 the attacker is trying to use it for accessing a file different from
 \texttt{evil\_file}. The \webapplication{} creates a nonce to ensure a
 fresh communication (42), performs a reading operation on the file-system
 (43--45), and sends an HTTP response to the client with the page
 \texttt{evil\_page} and the content of the file just read
 (\texttt{file(Path)}) (46).
 
 \begin{lstlisting}[breaklines=true, caption={ASLan++ code for the \chained{} case study.},label={lst:webappchained}]
 symbols
  Id: text;
  Username, Password, AuthCookie, Path, Name, Surname, Phone, Avatar: message;
  IP: agent;
  SQLquery, SQLresponse, Search: message;
  NonceWA : text;
 body{
 while(true){
 select{
  on( ? *->* Actor: ?IP.http_request(index,usr.s.?Username.s.pwd.s.?Password, none).tag1 ):{
  NonceWA := fresh();
  SQLquery := Username.Password;
  Actor *->* Database : query(SQLquery).NonceWA;
  select{
   on(Database *->* Actor : tuple(?SQLresponse).NonceWA):{
   AuthCookie := fresh();
   sessionValue(AuthCookie);
   Actor *->* IP: http_response(dashboard, tuple(SQLresponse).AuthCookie);
 }}}}}
  on( ? *->* Actor: ?IP.http_request(profileId, id.s.?Id, ?AuthCookie).tag3 ): {
   select{ on(sessionValue(AuthCookie)):{
    SQLquery := Id;
    NonceWA := fresh();
    Actor *->* Database : query(SQLquery).NonceWA;
   select{ on(Database *->* Actor : tuple(SQLresponse)):{
    Actor *->* IP : http_response(profileId,tuple(SQLresponse).AuthCookie);
 }}}}}
 on( ? *->* Actor : ?IP.http_request(edit, name.s.?Name.s.surname.s.?Surname.s.phone.s.?Phone.s.avatar.s.?Avatar,?AuthCookie).tag2):{
  select{ on(sessionValue(AuthCookie)):{
   SQLquery := Name.Surname.Phone.Avatar;
   NonceWA := fresh();
   Actor *->* Database : query(SQLquery).NonceWA;
   select{ on(Database *->* Actor : tuple(?SQLresponse).NonceWA):{
    NonceWA := fresh();
    Actor *->* Filesystem : writeFile(Avatar).NonceWA;
    Actor *->* IP : http_response(profileId, tuple(SQLresponse).AuthCookie);
 }}}}}
 % this branch represents an uploaded server-side code 
 % that reads from the filesystem
 on( ? *->* Actor : ?IP.http_request(evil_file, file.s.?Path, none).tag4):{
  select{on( inFS(evil_file) & evil_file != Path):{
  NonceWA := fresh();
  Actor *->* Filesystem : readFile(Path).NonceWA;
  %assert b:false;
  Filesystem *->* Actor : file(Path).NonceWA;
  Actor *->* IP : http_response(evil_file, file(Path));
 }}}}
 \end{lstlisting}
 
 As goal, once again, we use the same goal we defined in
 \autoref{lst:goal:fileInclusion} that states the attacker should not be
 able to access something which is function of \texttt{file()}.
 
 
\section{Concretization file}
\label{appendix:concretization}

 Along with the ASLan++ model, the security analyst has to define a
 concretization file that is used by \tool{} to carry out an automatic analysis.
 The concretization file is written in the JSON format
 and consists of the 
 information:
 \begin{compactitem}
 \item \texttt{domain}: the IP address of the server hosting the \webapplication{}.
 \item \texttt{tag\#}: The ASLan++ model will have a tag value
 \texttt{tag\#} associated to each HTTP request, where \texttt{\#} is an
 integer number. This tag value is used to map an abstract request in the
 ASLan++ file to the concretization details in the concretization file.
 Each \texttt{tag\#} defines a JSON object with the following information:
 \begin{compactitem}
 \item \texttt{url}: the URL of the page.
 \item \texttt{method}: whether the request to be performed is
 \texttt{GET}, \texttt{POST}, \texttt{PUT} or \texttt{CREATE}.
 \item \texttt{mapping}: a JSON dictionary that maps abstract parameters keys in the ASLan++ model to the real parameters keys.
 \item \texttt{params}: a JSON dictionary that defines the parameters
 required for performing the request. We use the placeholder \texttt{?} to
 define that a value of a parameter has to be decided at runtime, and
 \texttt{\_} to define that the value for that parameter is a file.
 \item \texttt{cookies}: a JSON dictionary that defines the cookies
 required for performing the request. We use \texttt{?} to
 define that a value of a cookie has to be decided at runtime.
 \item \texttt{tables}: a JSON dictionary that maps real parameters keys to
 the corresponding columns in the database in the form
 \texttt{table.column}.
 \end{compactitem}
 \item \texttt{abstract\_page}: for each abstract page represented in the
 ASLan++ model, we define the pair \texttt{abstract\_page:content\_to\_check},
 where \texttt{content\_to\_check} is a string that \tool{} will verify to be part of \texttt{abstract\_page} in order to ensure the page has been
 correctly retrieved.
 \item \texttt{evil\_file}: defines the local location of the malicious server side script that needs to be used.
 \end{compactitem}
 
 The concretization file for the \chained{} case study is reported in
 \autoref{lst:concretization:chained}.

 \begin{lstlisting}[breaklines=true,caption={
 	%JSON structure of the concretization 
 	Concretization file for the	\chained{} case study.},
 		label={lst:concretization:chained}]
 {"domain" : "127.0.0.1",
 "tag1" : {
 "url" : "https://127.0.0.1/index.php",
 "method" : "POST",
 "mapping" : { "usr" : "username" , "pwd" : "password" },
 "params" : { "username":"?" , "password":"?" },
 "tables" : {"username":"users.username", "password":"users.password"}},
 "tag2" : {
 "url" : "https://127.0.0.1/profile.php",
 "method" : "POST",
 "params" : { "avatar" : "_", "phone" : "?","surname":"?", "name":"?" },
 "mapping" : { "avatar" : "avatar" },
 "tables" : {"name":"users.name","surname":"users.surname","phone":"users.phone","avatar":"users.avatar"}},
 "tag3" : {
 "url" : "https://127.0.0.1/profile.php",
 "method" : "GET",
 "params" : { "id" : "?"},
 "mapping" : { "id" ;"id" },
 "tables" : {"id":"users.id"}},
 "tag4" : {
 "url" : "https://127.0.0.1/index.php",
 "method" : "GET",
 "params" : { "file" : "_" },
 "mapping" : { "file" : "file" }},
 "dashboard" : "Welcome",
 "profileId" : "Welcome",
 "files" :{
 "evil_file" : "/home/evil_file.txt"}}
 \end{lstlisting}

\end{EXT}

\end{document}